\documentclass[12pt]{iopart}

\usepackage{iopams}
\usepackage{graphicx}
\usepackage{bm}

\begin{document}

\title{Phase transitions of the Kane-Mele-Hubbard model with a long-range hopping}

\author{Tao Du, Yue-Xun Li, He-Lin Lu and Hui Zhang}

\address{Department of Physics, Yunnan Minzu University, Kunming 650504, P. R. China}
\ead{dutao@ymu.edu.cn}
\vspace{10pt}
\begin{indented}
\item[\today]
\end{indented}

\begin{abstract}
The interacting Kane-Mele model with a long-range hopping is studied using analytical method. The original Kane-Mele model is defined on a honeycomb lattice. In the work, we introduce a four-lattice-constant range hopping and the on-site Hubbard interaction into the model and keep its lattice structure unchanged. From the single-particle energy spectrum, we obtain the critical strength of the long-range hopping $t_L$ at which the topological transition occurs in the non-interacting limit of the model and our results show that it is independent of the spin-orbit coupling. After introducing the Hubbard interaction, we investigate the Mott transition and the magnetic transition of the generalized strongly correlated Kane-Mele model using the slave-rotor mean field theory and Hartree-Fock mean field theory respectively. In the small long-range hopping region, it is a correlated quantum spin Hall state below the Mott transition, while a topological Mott insulator above the Mott transition.  By comparing the energy band of spin degree of freedom with the one of electrons in non-interacting limit, we find a condition for the $t_L$-driven topological transition. Under the condition, critical values of $t_L$ at which the topological transition occurs are obtained numerically from seven self-consistency equations in both regions below and above the Mott transition. Influences of the interaction and the spin-orbit coupling on the topological transition are discussed in this work. Finally, we show complete phase diagrams of the generalized interacting topological model at some strength of spin-orbital coupling.
\end{abstract}

%\maketitle

\maketitle

\section{Introduction}\label{sec1}

Over the past decade, the research of topological insulators has been a main topic in condensed matter physics\cite{2010Hasan,2011Qi,2013Ando}. Topological insulators are novel quantum phases which have a charge excitation gap in bulk similarly to the band insulator and gapless edge modes described by Dirac-type Fermions. The topological phases are charactered by non-trivial topological invariants, e.g. TKNN number \cite{1982Thouless} or Chern number \cite{1988Haldane} and Z$_2$ topological invariant \cite{2005aKane}. The topological invariant should be changed when the topological phase transition occurs, which corresponds to the closing of the bulk gap due to the adiabatic continuity. By the argument of topology, it is believed that a topological insulator should be stable to weak disorder or many-body interactions as long as they keep the bulk gap opened \cite{2005aKane,1984Niu,1985Niu}. It is natural to investigate effects of strong electron-electron  correlations on topological systems and especilly search for new states of matter which arise due to the interplay between topology and strong correlations. The topological Mott insulator (TMI) \cite{2010Pesin} arising in correlated topological systems is a typical new state which has characters of Mott insulators (MI) and topological band insulators (TBI). Besides the TMI, a large number of topological phases emerge in various topologically non-trivial systems with strong correlations, such as the correlated topological insulator (CTI), the fractionalized quantum spin Hall state (FQSH), the fractionalized Chern insulator (FCI), etc. \cite{2015Neupert,2015Maciejko} These novel phases enrich the phase diagram of a topological system and phase transitions between them have attracted great attention over the past years\cite{2013Hohenadler,2014Imada,2014Witczak-Krempa}.

The Kane-Mele (KM) model \cite{2005bKane} is a significant toy model in studies of topological insulators. In the case of half-filling, the quantum spin Hall (QSH) state arises due to the spin-orbit coupling of next-nearest neighbor electrons on the honeycomb lattice. The QSH state characterized by a non-trivial Z$_2$ topological invariant has a bulk gap and an odd number of Kramers pairs of gapless edge modes \cite{2005aKane,2005bKane}. On the other hand, the Hubbard model \cite{1963Hubbard} may be the simplest possible model that captures the essential physics of strongly correlated systems, e.g. metal-insulator transitions. It is well known that there are two descriptions of metal-insulator transitions, i.e. the Mott scenario \cite{1963Hubbard,1964Hubbard} and the magnetic scenario \cite{1998Imada}. The combination of the two celebrated models, named Kane-Mele-Hubbard model(KMH), provides an ideal setting to reveal new physics of corrlated topological systems \cite{2008Young,2010Rachel,2012Ruegg,2012Vaezi,2012Wu,2011Yu,2011Hohenadler,2011Zheng,2012Hohenadler,2012Griset,2014Bercx}. Effects of strong electron correlations on the KM model on the honeycomb lattice have been investigated by various analytical or numerical methods, e.g. slave-particle/spin mean field methods \cite{2008Young,2010Rachel,2012Ruegg}, Schwinger boson/fermion approaches\cite{2012Vaezi}, the cellular dynamical mean field theory (DMFT) \cite{2012Wu}, the variational cluster approach (VCA) \cite{2011Yu} and the quantum Monte Carlo (QMC) simulation \cite{2011Hohenadler,2011Zheng,2012Hohenadler}. In general, the correlated QSH state of the KMH model which connects adiabatically to the QSH state of KM model is stable against the weaker correlation and the magnetic insulating phase emerges when the correlation becomes sufficiently strong. Between the two states various exotic states which stem from the interplay of topology and correlations, e.g. the TMI, the quantum spin liquid (QSL), and the QSH state coupled to a dynamical Z$_{2}$ gauge field (QSH$^{*}$), can emerge in the intermediate correlation region. Some of the phase transitions in KMH model have been investigated by Hohenadler \textit{et al.} \cite{2012Hohenadler} using the QMC simulation and Griset and Xu \cite{2012Griset} from the viewpoint of field theory. Furthermore, Bercx \textit{et al.} \cite{2014Bercx} have investigated effects of strong correlations on the KM model on the honeycomb lattice with a magnetic flux of $\pm \pi$ through each hexagon. They found that the antiferromagnetic order develops above a critical value of the correlation, similar to the case of ordinary KMH model, and there is a correlation-induced gap in the edge states as a result of umklapp scattering at half-filling. For comprehensive understanding on the field, we refer to recent review articles \cite{2013Hohenadler,2014Imada,2018Rachel}.

Recently, effects of interactions on various generalized KM models have been investigated by several anthors. Hung \textit{et al.} \cite{2013Hung,2014Hung} have studied a KMH model on the
honeycomb lattice with third-nearest neighbor hopping using the QMC simulation. Chen \textit{et al.} \cite{2015Chen} studied the same model on the same lattice at finite temperature using cellular dynamical mean field theory(DMFT). Based on the single-particle Green's function obtained from QMC calculations, Lang \textit{et al.} \cite{2013Lang} have investigated phase transitions driven by the bond dimerization and strong electron-electron correlations in the dimerized KMH model on the honeycomb lattice, such as the topological transition, the magnetic transition and the transition from non-ordered dimerized insulators to antiferromagnetic insulators. At the mean field level, Lai and Hung  \cite{2014Lai and Hung} have studied effects of the short-ranged interaction on the KM model with staggered potentials. They found that the on-site repulsive Hubbard interaction stabilizes the QSH state against the staggered potential, while the attractive interaction destabilizes the topological phase.

In the work of Hung \textit{et al.} \cite{2014Hung}, a KMH model with a long-range hopping (i.e. a four-lattice-constant range hopping) was investigated by the QMC simulation. In the non-interacting KM model, the long-range hopping can drive a transition from the Z$_2$ topological insulator to a topologically trivial band insulator. Their investigation showed that the on-site Hubbard interaction shifts the critical strength of the long-range hopping at which the topological phase transition occurs and the interaction stabilizes the topological insulator state against the long-rang hopping. It is necessary to note that different investigations between the two generalized KMH models \cite{2014Hung} with the third-nearest neighbor hopping and the long-range hopping stem from the type of hopping, since the two models have the same honeycomb lattice structure. From the discussion of a similar model in their work, such conclusion may be appropriate that the influence of the interaction on the critical strength of the long-range hopping can't be captured by Hartree-Fock (HF) mean field theory. 

In the present work, using mean field methods, we study the generalized KMH model which have been studied by Hung \textit{et al.}. In the case of non-interacting electrons, the topological phase transition driven by the four-lattice-constant range hopping is obtained from the single-particle energy spectrum. Slave-rotor mean field method \cite{2002Florens,2003Florens,2004Florens} is applied to the model when the Hubbard interaction is introduced. We show the interesting physics of KM model stemming from the interplay of the long-range hopping and strong interactions. It is also our concern that can the influence of the interaction on the topological transition driven by the long-range hopping be captured by slave-rotor mean field theory. Furthermore, HF mean field theory is applied to the large Hubbard interaction to investigate the magnetic transition of the generalized KMH model which have not been dealt with in other articles. The complete phase diagram including the correlated QSH state and the TMI is obtained in this work.

Our paper is organized as follows. In section 2, we revisit the KM model with the four-lattice-constant range hopping. The hopping strength at which the topological transition occurs is critical to the following discussion about the generalized interacting KM model. In section 3, the Hubbard interaction is introduced into the model and the two scenarios of metal-insulator transitions are obtained by slave-rotor mean field method and HF mean field theory respectively. In this section, we investigate in detail the topological phase transition driven by the long-range hopping and Hubbard interaction. Finally, we conclude in section 4.

\section{\label{sec2}The Kane-Mele model with the four-lattice-constant range hopping}

\subsection{\label{sec2.1}The model}

\begin{figure}[h]
\centering
\includegraphics[width=8.5cm,height=4.8cm]{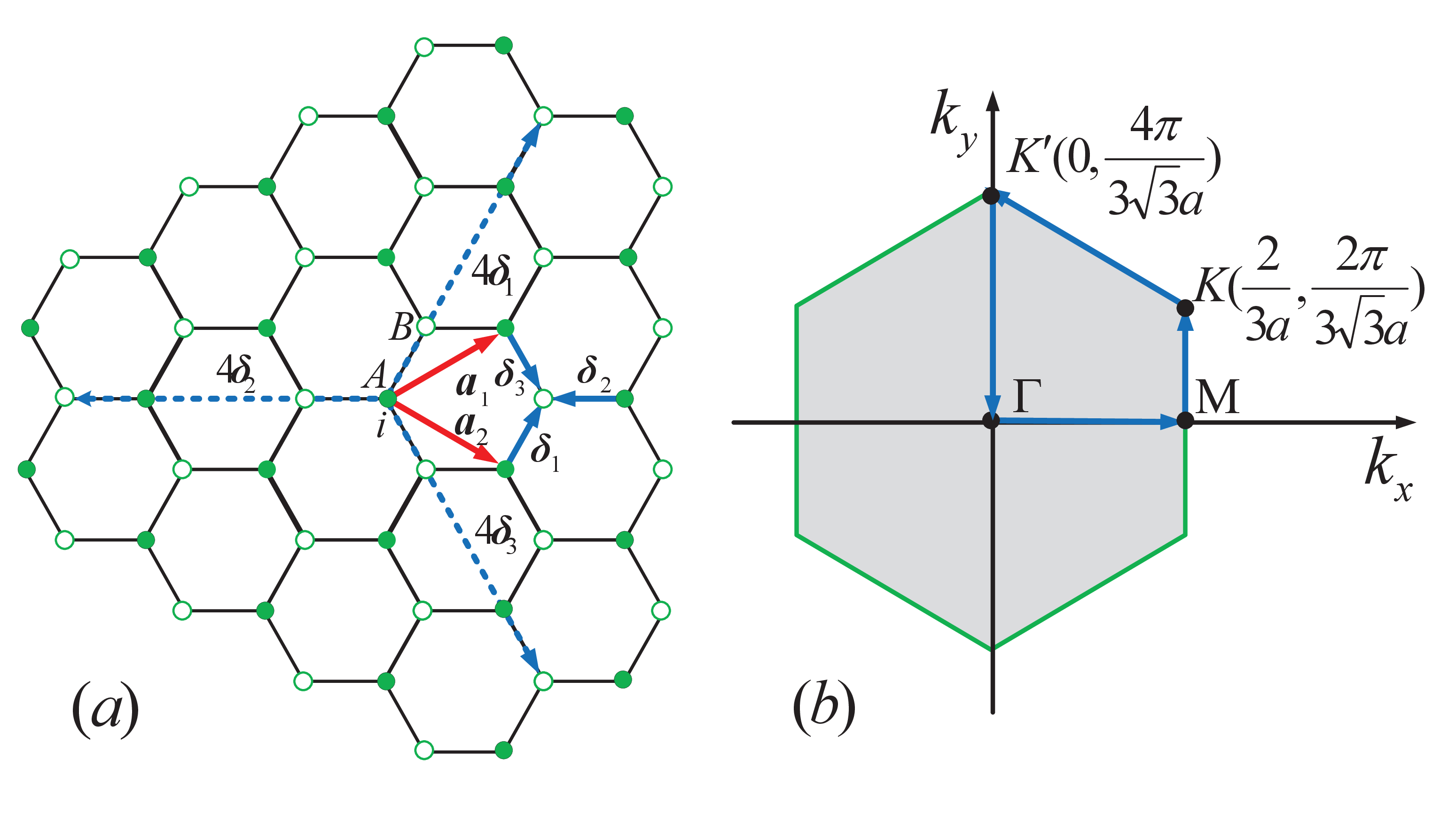}
\caption{\label{fig1}(a) the honeycomb lattice. Red solid arrows represent the lattice vectors ${\bm a}_{1}=(3a/2, \sqrt{3}a/2)$ and ${\bm a}_{2}=(3a/2, -\sqrt{3}a/2)$. Blue solid arrows represent the NN bonds in three directions: ${\bm \delta}_{1}=(a/2, \sqrt{3}a/2)$, ${\bm \delta}_{2}=(a/2, -\sqrt{3}a/2)$ and ${\bm \delta}_{3}=(-a, 0)$. The four-lattice-constant range bonds: $4{\bm \delta}_{1}$, $4{\bm \delta}_{2}$ and $4{\bm \delta}_{3}$ are represented by blue dotted arrows. (b) The Brillouin zone of the model.}
\end{figure}

The generalized Kane-Mele model on the honeycomb lattice is

\begin{equation}
H_{0}=-t\sum_{<ij>}\sum_{\sigma}\hat{c}^{\dagger}_{i\sigma}\hat{c}_{j\sigma}-t_{L}\sum_{\{ij\}}\sum_{\sigma}\hat{c}^{\dagger}_{i\sigma}\hat{c}_{j\sigma}+\mathrm{i}\lambda\sum_{\ll ij\gg}\sum_{\sigma\sigma^{\prime}}\nu_{ij}\hat{c}_{i\sigma}^{\dagger}\sigma_{\sigma\sigma^{\prime}}^{z}\hat{c}_{j\sigma^{\prime}}.
\label{eq1}
\end{equation}
Here $\hat{c}_{i\sigma}^{\dagger}$ ($\hat{c}_{i\sigma}$) is an creation (annihilation) operator of an electron with spin $\sigma=\pm1$ at site $i$, $\sigma_{\sigma\sigma^{'}}^{z}$ is the $z$ component of Pauli matrices, $t$ is the hopping strength of nearest neighbor (NN) electrons, and $\lambda$ is the strength of spin-orbit coupling of next-nearest neighbor (NNN) electrons. The second term is the four-lattice-constant range hopping term with the strength $t_{L}$. The parameter $\nu_{ij}=-1$ if the orientation of the NNN sites $i$, $j$ is right turn while $\nu_{ij}=+1$ if left turn. Lattice vectors of the honeycomb lattice are ${\bm a}_{1}=(3a/2, \sqrt{3}a/2)$ and ${\bm a}_{2}=(3a/2, -\sqrt{3}a/2)$, as shown in figure~\ref{fig1}. In the work, we set the lattice constant $a=1$ and the strength of NN hopping $t=1$.

In momentum space, the Hamiltonian of the so called $t_L$-KM model can be obtained as 
\begin{equation}
H_{0}=\sum_{\bm k}\Psi_{\bm k}^{\dagger}{\cal H}_{0\bm k}\Psi_{\bm k}.\label{eq2}
\end{equation}
Here $\Psi_{\bm k}=(\hat{c}_{{\bm k}\uparrow}^A, \hat{c}_{{\bm k}\uparrow}^B, \hat{c}_{{\bm k}\downarrow}^A, \hat{c}_{{\bm k}\downarrow}^B)^T $ is the electron operator in the momentum-spin space and the Bloch Hamiltonian ${\cal H}_{0\bm k}$ is
\begin{eqnarray}
\left(
\begin{array}{cccc}
\lambda\gamma&-g_{t_{L}}&0&0\\
-g_{t_{L}}^{*}&-\lambda\gamma&0&0\\
0&0& -\lambda\gamma&-g_{t_{L}}\\
0&0&-g_{t_{L}}^{*}&\lambda\gamma\\
\end{array}
\right),\nonumber
\end{eqnarray}
where A, B represent the sublattice of the honeycomb lattice as shown in figure~\ref{fig1},  $g_{t_{L}}=t\sum_{i}e^{\mathrm{i}{\bm k}\cdot\bm \delta_{i}}+t_{L}\sum_{i}e^{\mathrm{i}{\bm k}\cdot4\bm \delta_{i}}$, and  $\gamma=2[-\sin(\sqrt{3}k_{y})+2\cos(3k_x/2)\sin(\sqrt{3}k_y/2)$.
From the Hamiltonian matrix ${\cal H}_{0\bm k}$, the single-particle energy spectrum can be obtained as
\begin{equation}
E_{\pm}(\bm k)=\pm\sqrt{|g_{t_{L}}|^2+(\lambda\gamma)^2}
\label{eq3}
\end{equation}

\subsection{\label{sec2.2}The topological phase transition driven by the long-range hopping}

\begin{figure}[h]
\centering
\includegraphics[width=8.2cm,height=5.1cm]{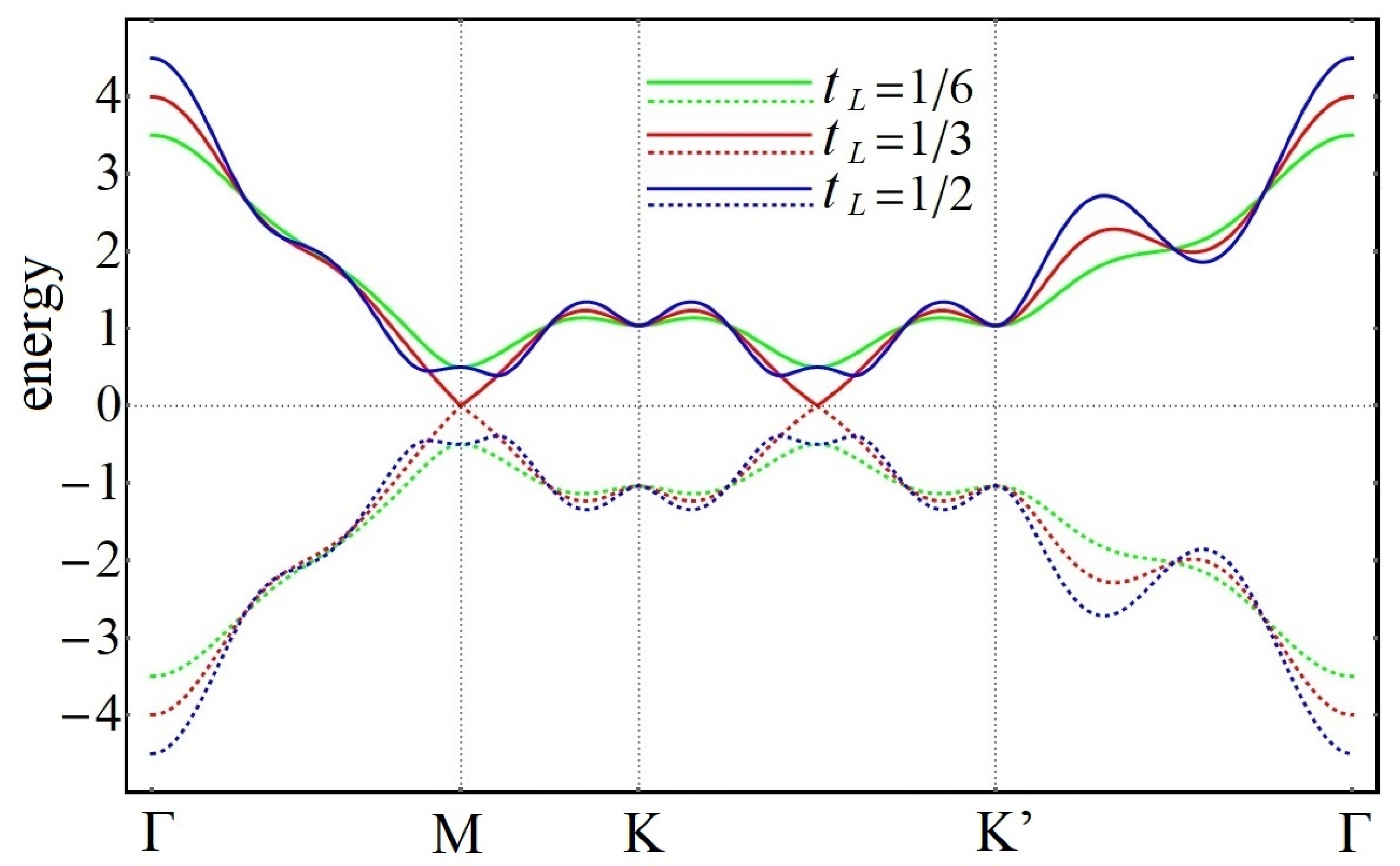}
\caption{\label{fig2}Energy band structure of equation~(\ref{eq3}) at the strength of spin-orbit coupling $\lambda=0.2$. The path in the Brillouin zone is taken as shown in the figure~\ref{fig1}(b).}
\end{figure}

\begin{figure}[h]
\centering
\includegraphics[width=8.2cm,height=5.1cm]{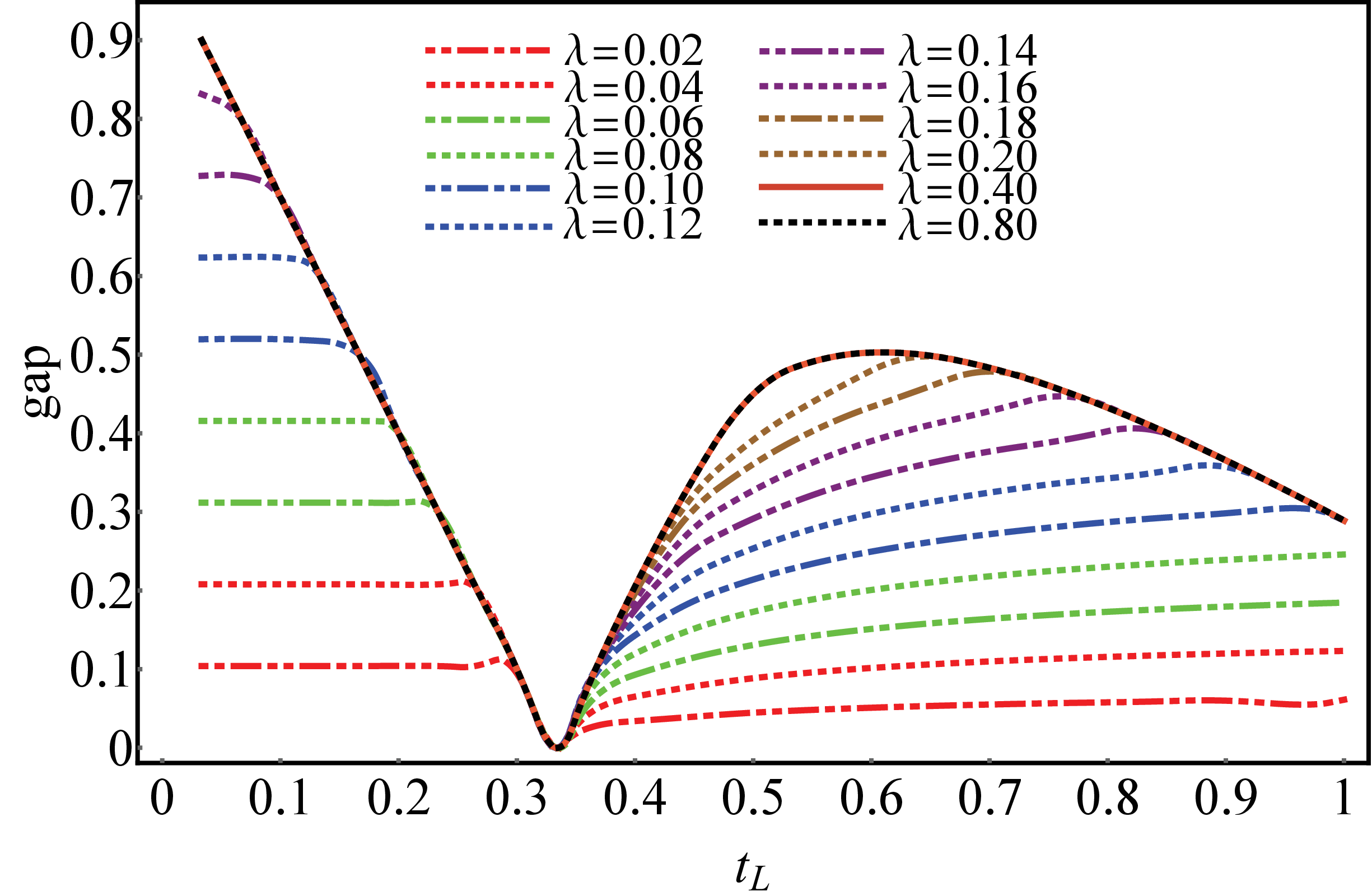}
\caption{\label{fig3}Gaps with the change of $t_{L}$ at various values of $\lambda$.}
\end{figure}

The energy spectum (equation~(\ref{eq3})) of the $t_L$-KM model becomes the one of KM model when the strength of the long-range hopping $t_{L}=0$. The KM model has a energy gap when $\lambda\not=0$ and possesses a QSH state at half-filling. Although the energy gap of equation~(\ref{eq3}) varies with the increase of the strength $t_{L}$, the $t_L$-KM model still stays in a QSH state which connects adiabatically to the topological state of KM model as long as $t_L$ keeps the gap opened . When the gap closes (then reopens) at some values of $t_{L}$, a topological transition from the QSH state to the topologically trivial band insulator occurs. At $\lambda=0.2$, the band structure for various $t_L$ is shown in figure~\ref{fig2}. It is obvious that gap is closed at $t_{L}=1/3$ and nodes are localized at the M-point and the midpoint of the K-point and K$^{\prime}$-point along the boundary of the Brillouin zone. Numerically, it is easy to find that gaps always close at $t_{L}=1/3$ and is independent of the strength of spin-orbital coupling. Figure~\ref{fig3} shows gaps of the generalied KM model for various values of $\lambda$. When $t_{L}>1/3$ and $\lambda\not=0$, the gap opens again and the phase is a topologically trivial band insulator.

\section{\label{sec3}The $t_L$-Kane-Mele model with strong interactions}

In the section, the on-site Hubbard interaction is introduced into the $t_L$-KM model (called $t_L$-KMH model). As mentioned earlier, there are two scenarios of metal-insulator transitions in systems with strong electron-electron correlations. In the Mott scenario, strongly correlated systems often
display the spin-charge separation. It postulates that electrons in those systems can be viewed as composites of chargons and spinons, which respectively describe the charge and spin degrees of freedom. The slave-rotor representation of the physical electron operators can treat economically the spin-charge separation \cite{2002Florens,2003Florens,2004Florens} and describe appropriately the Mott transition of the charge degree of freedom and the possible quantum spin liquid (QSL) ground state of the spin degree of freedom in strongly correlated systems\cite{2005Lee,2007Zhao,2008Senthil}. For the $t_L$-KMH model studied here, in the slave-rotor representation the decoupling of charge and spin degrees of freedom can lead to a novel state of matter---the QSL with non-trivial topological band structure. We capture the Mott transition of the charge degree of freedom at the intermediate Hubbard interaction using the slave-rotor mean field method. The boundary of the Mott transition is obtained numerically from the mean field self-consistency equations in the section. Furthermore, the slave-rotor mean field method can be applied to obtain $t_{L}$-driven topological transitions in both cases of condensed and uncondensed charges. The influences of the Hubbard interaction on $t_{L}$-driven topological transitions is also discussed in detail here.

In the ordinary Hubbard model on the honeycomb lattice, the magnetic order can emerge at the large Hubbard interaction, although the critical Hubbard interaction strongly depends on the used method, e.g. HF mean field theory, QMC simulation and dynamical mean field theory (DMFT)\cite{1996Sorella,2009Jafari}. Similar situations arise when we consider the $t_L$-KMH model at the large Hubbard interaction. In this section, we obtain the boundary of the magnetic transition by the HF mean field theory for simplicity and focus more on the Mott tansition and influences of the Hubbard interaction on $t_{L}$-driven topological transitions using the slave-rotor mean field method.     
\subsection{\label{sec3.1}The Mott transition of the charge degree of freedom}

\subsubsection{\label{sec3.1.1}Slave-rotor representation.}

In the slave-rotor representation\cite{2002Florens,2003Florens,2004Florens}, the electron annihilation operator is decomposed as
\begin{equation}
\hat{c}_{i\sigma}=e^{\mathrm{i}\theta_{i}}\hat{f}_{i\sigma}.
\label{eq4}
\end{equation}
Here $e^{\mathrm{i}\theta_{i}}$ is the U(1) rotor operator that describes the charge degree of freedom and  $\hat{f}_{i\sigma}$ is the spinon operator that describes the spin degree of freedom of the electron. To recover the Hilbert space of the electron, the charge and spin degree of freedom should satisfy the constraint
\begin{equation}
\sum_{\sigma}\hat{f}_{i\sigma}^{\dagger}\hat{f}_{i\sigma}+ \hat{L}_{i}=1.
\label{eq5}
\end{equation}
Where the canonical angular momentum $\hat{L}_{i}=\mathrm{i}\partial_{\theta_{i}}$ associated with the angular $\theta_{i}$ is introduced. 

When the on-site Hubbard term is introduced, the Hamiltonian of the $t_{L}$-KMH model reads
\begin{equation}
H=H_{0}+\frac{U}{2}\sum_{i}\Big(\sum_{\sigma}n_{i\sigma}-1\Big)^{2}.
\label{eq6}
\end{equation}
In the slave-rotor representation, the Hamiltonian can be written as
\begin{eqnarray}
H&=&-t\sum_{<ij>\sigma}e^{-\mathrm{i}\theta_{ij}}\hat{f}^{\dagger}_{i\sigma}\hat{f}_{j\sigma}-t_{L}\sum_{\{ij\}\sigma}e^{-\mathrm{i}\theta_{ij}}\hat{f}^{\dagger}_{i\sigma}\hat{f}_{j\sigma}\nonumber\\
&&+\mathrm{i}\lambda\sum_{\ll ij\gg}\sum_{\sigma\sigma^{\prime}}\nu_{ij}e^{-\mathrm{i}\theta_{ij}}\hat{f}_{i\sigma}^{\dagger}\sigma_{\sigma\sigma^{\prime}}^{z}\hat{f}_{j\sigma^{\prime}}\nonumber\\
&&+\frac{U}{2}\sum_{i}\hat{L}_{i}^{2}-\mu\sum_{i\sigma}\hat{f}_{i\sigma}^{\dagger}\hat{f}_{j\sigma}.
\label{eq7}
\end{eqnarray}
Here, $\theta_{ij}=\theta_{i}-\theta_{j}$, and $\mu$ is the chemical potential. From equation~(\ref{eq7}), it is clear that the spin degree of freedom has the same band structure as the electron in non-interacting limit when $\left<e^{-\mathrm{i}\theta_{ij}}\right>\not=0$ at the mean field level. The partition function is written as a path integral of $e^{-S_{E}}$ over fields $f$, $f^{*}$ and $\theta$, where
\begin{equation}
\fl S_{E}=\int_{0}^{\beta}d\tau\Big[\sum_{i}-\mathrm{i}L_{i}\partial_{\tau}\theta_{i}+\sum_{i\sigma}f^{*}_{i\sigma}\partial_{\tau}f_{i\sigma}+H\nonumber+\sum_{i}h_{i}\Big(\sum_{\sigma}f^{*}_{i\sigma}f_{i\sigma}+L_{i}-1\Big)\Big]
\label{eq8}
\end{equation}
is the action in \textit{imaginary time} ($\tau=\mathrm{i}t$). Here the Lagrange multiplier $h_{i}$ is introduced into the action to impose the constraint of equation~(\ref{eq5}). From the canonical equation of motion in imaginary time, i.e. $\mathrm{i}\partial_{\tau}\theta_{i}=\partial H/\partial L_{i}$, we can obtain the relation of $L$ and $\theta$ as $L_{i}=(\mathrm{i}/U)\partial_{\tau}\theta_{i}$. Therefore, the action of the $t_L$-KMH model can be given as

\begin{eqnarray}
S_{E}&=&\int_{0}^{\beta}d\tau\Big[\sum_{i\sigma}f^{*}_{i\sigma}(\partial_{\tau}-\mu+h_{i})f_{i\sigma}+\frac{1}{2U}\sum_{i}(\partial_{\tau}\theta_{i}+\mathrm{i}h_{i})^{2}\nonumber\\
&&+\sum_{i}(-h_{i}+\frac{h_{i}^{2}}{2U})-t\sum_{<ij>\sigma}e^{-\mathrm{i}\theta_{ij}}f^{*}_{i\sigma}f_{j\sigma}-t_{L}\sum_{\{ij\}\sigma}e^{-\mathrm{i}\theta_{ij}}f^{*}_{i\sigma}f_{j\sigma}\nonumber\\
&&+\mathrm{i}\lambda\sum_{\ll ij\gg}\sum_{\sigma\sigma^{\prime}}\nu_{ij}e^{-\mathrm{i}\theta_{ij}}f_{i\sigma}^{*}\sigma_{\sigma\sigma^{\prime}}^{z}f_{j\sigma^{\prime}}\Big].
\label{eq9}
\end{eqnarray}
Next, we introduce a new field $X_{i}=e^{\mathrm{i}\theta_{i}}$ which is imposed by the constraint $|X_{i}|^2=1$ and six mean field parameters as follows:
\begin{eqnarray}
Q_{X}=\big<\sum_{\sigma}f_{i\sigma}^{A*}f_{j\sigma}^{B}\big>_{<ij>},\\
\label{eq10}
Q_{f}=\big<e^{-\mathrm{i}\theta_{ij}}\big>_{<ij>},\\
\label{eq11}
Q_{X}^{\prime}=\big<\sum_{\sigma\sigma^{\prime}}\mathrm{i}\nu_{ij}f_{i\sigma}^{*}\sigma_{\sigma\sigma^{\prime}}^{z}f_{j\sigma^{\prime}}\big>_{\ll ij \gg},\\
\label{eq12}
Q_{f}^{\prime}=\big<e^{-\mathrm{i}\theta_{ij}}\big>_{\ll ij \gg},\\
\label{eq13}
Q_{X}^{\prime\prime}=\big<\sum_{\sigma}f_{i\sigma}^{A*}f_{j\sigma}^{B}\big>_{\{ij\}},\\
\label{eq14}
Q_{f}^{\prime\prime}=\big<e^{-\mathrm{i}\theta_{ij}}\big>_{\{ij\}}.
\label{eq15}
\end{eqnarray}
Then, the action becomes quadratic in the mean field treatment and is described by the following expression in terms of two degrees of freedom (i.e. the charge degree of freedom $X$ and spin degree of freedom $f$)
\begin{equation}
\fl S_{E}=\int_{0}^{\beta}d\tau\Big[\frac{1}{2U}\sum_{i}\mathrm{i}\partial_{\tau}X_{i}^{*}(-\mathrm{i}\partial_{\tau})X_{i}+\sum_{i}\rho_{i}|X_{i}|^{2}+H^{X}+\sum_{i\sigma}f_{i\sigma}^{*}\partial_{\tau}f_{i\sigma}+H^{f}+\cdots\Big].
\label{eq16}
\end{equation}
Here, the symbol $``\cdots"$ denotes constant terms of mean field decomposition and we have set $h_{i}\equiv h=-\mu=0$ for half-filling at the mean field level. $\rho_{i}$ is the Lagrange multiplier for constraint $|X_{i}|^2=1$ and $\rho_i\equiv\rho$ in the mean field treatment. In the above expression,
\begin{equation}
H^{X}=-tQ_{X}\sum_{<ij>}X_{i}^{*}X_{j}-t_{L}Q_{X}^{\prime\prime}\sum_{\{ij\}}X_{i}^{*}X_{j}\nonumber+\lambda Q_{X}^{\prime}\sum_{\ll ij \gg}X_{i}^{*}X_{j}
\label{eq17}\end{equation}and 
\begin{equation}
H^{f}=-tQ_{f}\sum_{<ij>\sigma}f_{i\sigma}^{*}f_{j\sigma}-t_{L}Q_{f}^{\prime\prime}\sum_{\{ij\}\sigma}f_{i\sigma}^{*}f_{j\sigma}+\mathrm{i}\lambda Q_{f}^{\prime}\sum_{\ll ij \gg}\sum_{\sigma\sigma^{\prime}}\nu_{ij}f_{i\sigma}^{*}\sigma_{\sigma\sigma^{\prime}}^{z}f_{j\sigma^{\prime}}.
\label{eq18}
\end{equation}
The action of equation~(\ref{eq16}) can be transformed into frequency-momentum space via Fourier transforms
\begin{eqnarray}
&&X_{i}(\tau)=\frac{1}{\sqrt{\beta N_{\Lambda}}}{\sum_{\bm{k},n}}'e^{\mathrm{i}(\bm{k}\cdot\bm{R}_{i}-v_{n}\tau)}X_{\bm{k}}(\mathrm{i}v_{n})+\sqrt{x_{0}}
\label{eq19},\\
&&f_{i\sigma}(\tau)=\frac{1}{\sqrt{\beta N_{\Lambda}}}{\sum_{\bm{k},n}}e^{\mathrm{i}(\bm{k}\cdot\bm{R}_{i}-\omega_{n}\tau)}f_{\bm{k}\sigma}(\mathrm{i}\omega_{n})
\label{eq20}.
\end{eqnarray}
Here $N_{\Lambda}$ denotes the number of unit cells and $x_{0}$ is the density of the condensate of charges. $v_{n}=2n\pi/\beta$ are the Matsubara frequencies for bosons and $\omega_{n}=(2n+1)\pi/\beta$ for fermions and the summation excludes the point $(\mathrm{i}v_{n}^0,\bm{k}^0)$ at which the condensate of charges occurs. Finally, we can write the action in matrix form as
\begin{equation}
\fl S_{E}=\sum_{\bm{k},n}\Psi_{\eta}^{X\dagger}\Big[\Big(\frac{v_{n}^{2}}{2U}+\rho\Big)\delta_{\eta\kappa}+{\cal{H}}_{\eta\kappa}^{X}\Big]\Psi_{\kappa}^{X}+\sum_{\bm{k},n}\Psi_{\eta}^{f\dagger}\big[\big(-\mathrm{i}\omega_{n}\big)\delta_{\eta\kappa}+{\cal{H}}_{\eta\kappa}^{f}\big]\Psi_{\kappa}^{f}+\cdots
\label{eq21}
\end{equation}
Here $\Psi^{X}=\big(X_{\bm{k}}^{A}(\mathrm{i}v_{n}),X_{\bm{k}}^{B}(\mathrm{i}v_{n})\big)^{T}$ and $\Psi^{f}=\big(f_{\bm{k}\uparrow}^{A}(\mathrm{i}\omega_{n}),$
$f_{\bm{k}\uparrow}^{B}(\mathrm{i}\omega_{n}),f_{\bm{k}\downarrow}^{A}(\mathrm{i}\omega_{n}),f_{\bm{k}\downarrow}^{B}(\mathrm{i}\omega_{n}\big)^{T}$. 
Hamiltonian matrices of the X-field and $f$-field are respectively
\begin{equation}
{\cal{H}}^{X}=
\left(
\begin{array}{cc}
Q_{X}^{\prime}\lambda \gamma_{X}&-Q_{X}g-Q_{X}^{\prime\prime}g_{t_{L}}^{\prime}\\
-Q_{X}g^{*}-Q_{X}^{\prime\prime}g_{t_{L}}^{\prime*}&Q_{X}^{\prime}\lambda \gamma_{X}\\
\end{array}
\right)
\end{equation}\label{eq22}
and
\begin{equation}
\fl {\cal{H}}^{f}=
\left(
\begin{array}{cccc}
Q_{f}^{\prime}\lambda \gamma_{f}&-Q_{f}g-Q_{f}^{\prime\prime}g_{t_{L}}^{\prime}&0&0\\
-Q_{f}g^{*}-Q_{f}^{\prime\prime}g_{t_{L}}^{\prime*}&-Q_{f}^{\prime}\lambda \gamma_{f}&0&0\\
0&0&-Q_{f}^{\prime}\lambda \gamma_{f}&-Q_{f}g-Q_{f}^{\prime\prime}g_{t_{L}}^{\prime}\\
0&0&-Q_{f}g^{*}-Q_{f}^{\prime\prime}g_{t_{L}}^{\prime*}&Q_{f}^{\prime}\lambda \gamma_{f}\\
\end{array}
\right)
\end{equation}\label{eq23}
Here $\gamma_{X}=2[\cos(\sqrt{3}k_{y})+2\cos(3k_{x}/2)\cos(\sqrt{3}k_{y}/2)]$,  $\gamma_{f}=\gamma=2[-\sin(\sqrt{3}k_{y})+2\cos(3k_x/2)\sin(\sqrt{3}k_y/2)$, $g=t\sum_{i}e^{\mathrm{i}{\bm k}\cdot{\bm \delta}_{i}}$, and $g_{t_{L}}^{\prime}=t_{L}\sum_{i}e^{\mathrm{i}{\bm k}\cdot4\bm \delta_{i}}$.

\subsubsection{\label{sec3.1.2}Green's functions of two degrees of freedom and self-consistency equations for the Mott transition.}

From the action of equation~(\ref{eq21}), Green's functions of the charge degree of freedom and spin degree of freedom in the lower band are respectively\cite{2015Coleman}
\begin{equation}
G_{X}^{l}=\frac{1}{v_{n}^{2}/U+\rho+E_{-}^{X}}
\label{eq24}
\end{equation}and
\begin{equation}
G_{f\uparrow(\downarrow)}^{l}=\frac{1}{\mathrm{i}\omega_{n}-E_{-}^{f}}.
\label{eq25}
\end{equation}
Here $E_{-}^{X}=-|Q_{X}g+Q_{X}^{\prime\prime}g_{t_{L}}^{\prime}|+Q_{X}^{\prime}\lambda\gamma_{X}$ and $E_{-}^{f}=-\sqrt{|Q_{f}g+Q_{f}^{\prime\prime}g_{t_{L}}^{\prime}|^{2}+\big(Q_{f}^{\prime}\lambda\gamma_{f})^{2}}$ are respectively lower energy eigenvalues of Hamiltonian matrices ${\cal{H}}^{X}$ and ${\cal{H}}^{f}$. In the expression of equation~(\ref{eq24}), we have replaced $U$ by $U/2$ to preserve the exact atomic limit\cite{2002Florens,2004Florens}. In fact, energy spectrums of the lower band of charge and spin degree of freedom can be obtained respectively from Green's functions as
\begin{equation}
\xi^{l}(\bm{k})=\sqrt{U(\rho+E_{-}^{X})}
\label{eq26}
\end{equation}and
\begin{equation}
\Xi^{l}(\bm{k})=E_{-}^{f}.
\label{eq27}
\end{equation}
It is noteworthy that the energy spectrum $\xi^{l}(\bm{k})$ of the lower band of charge degree of freedom is different from the lower energy eigenvalue $E_{-}^{X}$ of the Hamiltonian matrice ${\cal{H}}^{X}$.

The six mean field parameters have been introduced in the slave-rotor representation, i.e. equation~(\ref{eq10})--~(\ref{eq15}). Including the constraint equation of the X-field (i.e. $|X_{i}|^{2}=1$), there are seven self-consistency equations in the slave-rotor mean field method. When the Mott transition occurs, the density of the condensate of charges $x_0=0$ and $\rho=-\mathrm{min}(E_{-}^{X})$ derived from equation~(\ref{eq26}). Then the self-consistency equations for the Mott transition of the charge degree of freedom can be obtained as follows
\begin{eqnarray}
\frac{1}{N_{\Lambda}}\sum_{\bm{k}}\frac{\sqrt{U_{c}}}{2\sqrt{E_{-}^{X}-\mathrm{min}(E_{-}^{X})}}=1
\label{eq28}\\
Q_{X}=\frac{1}{6N_{\Lambda}t}\sum_{\bm{k}}\frac{2Q_{f}|g|^{2}+Q_{f}^{\prime\prime}(gg_{t_{L}}^{\prime*}+g_{t_{L}}^{\prime}g^{*})}{\sqrt{(Q_{f}^{\prime}\lambda\gamma_{f})^{2}+|Q_{f}g+Q_{f}^{\prime\prime}g_{t_{L}}^{\prime}|^{2}}}
\label{eq29}\\
Q_{X}^{\prime}=\frac{1}{6N_{\Lambda}}\sum_{\bm{k}}\frac{-Q_{f}^{\prime}\lambda\gamma_{f}^{2}}{\sqrt{(Q_{f}^{\prime}\lambda\gamma_{f})^{2}+|Q_{f}g+Q_{f}^{\prime\prime}g_{t_{L}}^{\prime}|^{2}}}
\label{eq30}\\
Q_{X}^{\prime\prime}=\frac{1}{6N_{\Lambda}t_{L}}\sum_{\bm{k}}\frac{Q_{f}^{\prime}(gg_{t_{L}}^{\prime*}+g_{t_{L}}^{\prime}g^{*})+2Q_{f}^{\prime\prime}|g_{t_{L}}^{\prime}|^{2}}{\sqrt{(Q_{f}^{\prime}\lambda\gamma_{f})^{2}+|Q_{f}g+Q_{f}^{\prime\prime}g_{t_{L}}^{\prime}|^{2}}}
\label{eq31}\\
Q_{f}=\frac{1}{12N_{\Lambda}t}\sum_{\bm{k}}\frac{\sqrt{U_{c}}[2Q_{X}|g|^{2}+Q_{X}^{\prime\prime}(gg_{t_{L}}^{\prime*}+g_{t_{L}}^{\prime}g^{*})]}{2\sqrt{E_{-}^{X}-\mathrm{min}(E_{-}^{X})}|Q_{X}g+Q_{X}^{\prime\prime}g_{t_{L}}^{\prime}|}
\label{eq32}\\
Q_{f}^{\prime}=\frac{1}{12N_{\Lambda}}\sum_{\bm{k}}\frac{\sqrt{U_{c}}\gamma_{X}}{2\sqrt{E_{-}^{X}-\mathrm{min}(E_{-}^{X})}}
\label{eq33}\\
Q_{f}^{\prime\prime}=\frac{1}{12N_{\Lambda}t_{L}}\sum_{\bm{k}}\frac{\sqrt{U_{c}}[Q_{X}(gg_{t_{L}}^{\prime*}+g_{t_{L}}^{\prime}g^{*})+Q_{X}^{\prime\prime}|g_{t_{L}}^{\prime}|^{2}]}{2\sqrt{E_{-}^{X}-\mathrm{min}(E_{-}^{X})}|Q_{X}g+Q_{X}^{\prime\prime}g_{t_{L}}^{\prime}|}.
\label{eq34}
\end{eqnarray}
Here $U_{c}$ is the critical Hubbard interaction at which the Mott transition of the charge degree of freedom occurs.

\subsubsection{\label{sec3.1.3}Results.}

\begin{figure}[h]
\centering
\includegraphics[width=8.2cm,height=5.1cm]{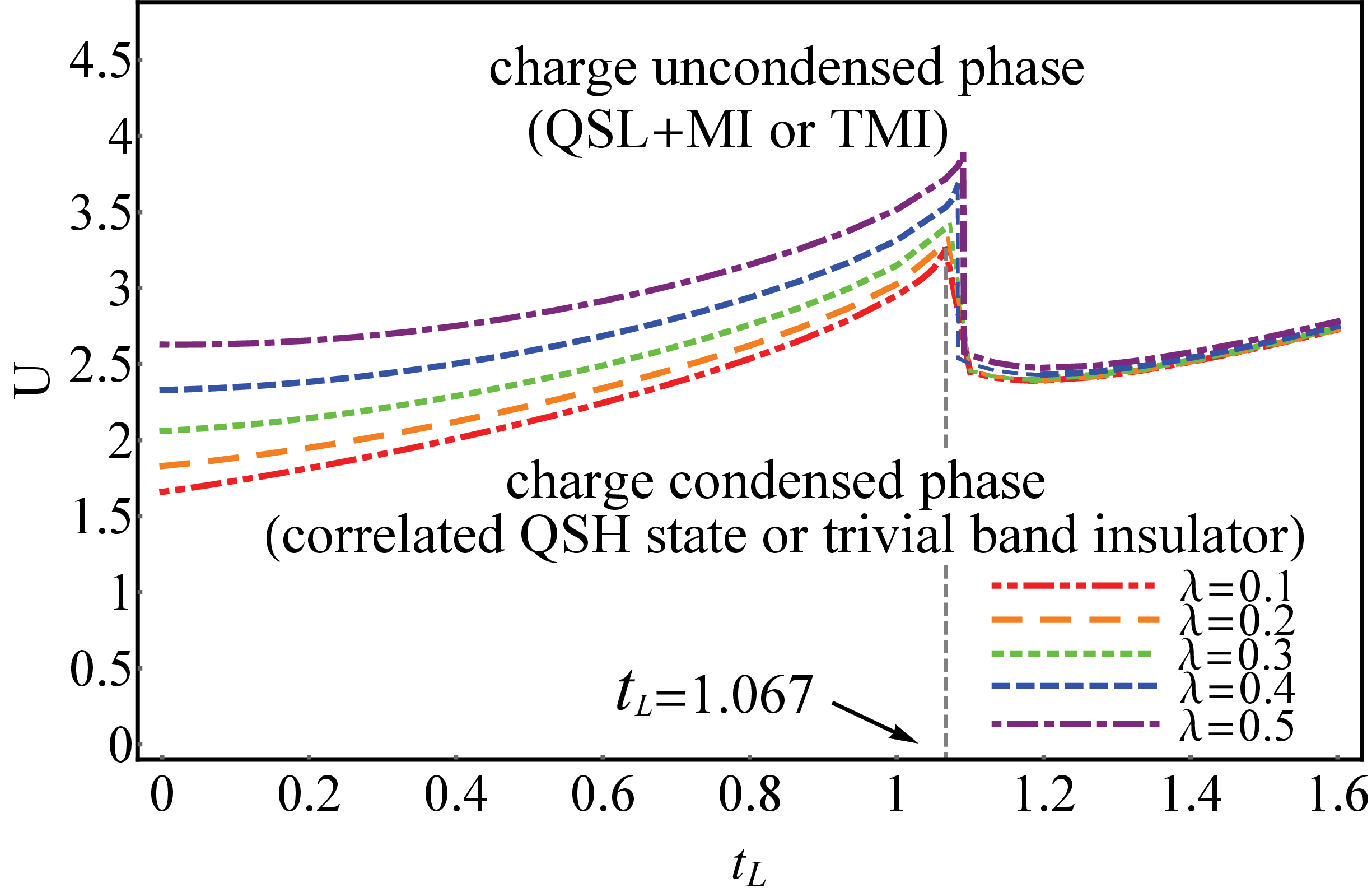}
\caption{\label{fig4}Mott transitions of the charge degree of freedom. QSH: quantum spin Hall, QSL: quantum spin liquid, MI: Mott insulator, and TMI: topological Mott insulator.}
\end{figure}

Solving numerically the seven self-consistency equations, i.e. equation~(\ref{eq28})--(\ref{eq34}), we obtain boundaries of the Mott transition for various $\lambda$ as shown in figure~\ref{fig4}.
In the small-U region, the gap of the charge degree of freedom is closed and charges are condensed, i.e. $\left<X_{i}\right>\not=0$. In this case, the charge degree of freedom combines the spin degree of freedom to form the conventional electron with the same band structure as the electron in $t_{L}$-KM model (see equation(\ref{eq7}) or (\ref{eq18})). When the spin degree of freedom has a topologically non-trivial band structure, the combined phase is a \textit{correlated} QSH state which connect adiabatically to the QSH state possessed by the $t_{L}$-KM model. It is clear that the long-range hopping can drive the topological transition from QSH insulators to topologically trivial band insulators. Therefore, the charge condensed phase may be a trivial band insulator when the strength  $t_{L}$ of the long-range hopping beyond some critical values at which the gap of the spin degree of freedom closes (then reopens).

In the larger-U region, the gap of the charge degree of freedom opens and there is a spin-charge separation. It is a Mott insulator (MI) for the charge degree of freedom, while a quantum spin liquid (QSL) state for the spin degree of freedom. When the spin degree of freedom has a topologically non-trivial band structure, the charge uncondensed phase is a novel topological Mott insulator (TMI)\cite{2010Pesin}. As the case of condensed charges, there may be a $t_{L}$-driven topological transition. If it occurs, the charge uncondensed phase is just a mixing phase of the MI and 
a conventional QSL state.

The $t_L$-driven topological transitions in both regions will be discussed in section~\ref{sec3.3} and ~\ref{sec3.4}.

\subsection{\label{sec3.2}The magnetic transition}

In this subsection, we obtain the boundary of the magnetic transition using HF mean field theory. A spin density wave (SDW) phase can emerge in various Hubbard models on the honeycomb lattice, regardless of whether there is a topological band structure or not \cite{2010Rachel,1996Sorella,2009Jafari,2011He,2016Li}. In our KMH model, the topologically non-trivial band structure of the spin degree of freedom should be destroyed due to the breaking of time-reversal symmetries with the development of magnetic order. In the case of large on-site Hubbard interactions, the charge uncondensed phase observed in the last subsection inevitably turns into a SDW phase due to the bipartite nature of the honeycomb lattice.

The Hubbard interaction is 
\begin{eqnarray}
H_{U}&=&U\sum_{i=1}^{2N_{\Lambda}}n_{i\uparrow}n_{i\downarrow}\nonumber\\
&=&\frac{U}{4}\sum_{i=1}^{2N_{\Lambda}}\left[(n_{i\uparrow}+n_{i\downarrow})^{2}-(n_{i\uparrow}-
n_{i\downarrow})^{2}\right].
\label{eq35}
\end{eqnarray}

\begin{figure}[h]
\centering
\includegraphics[width=8.2cm,height=5.2cm]{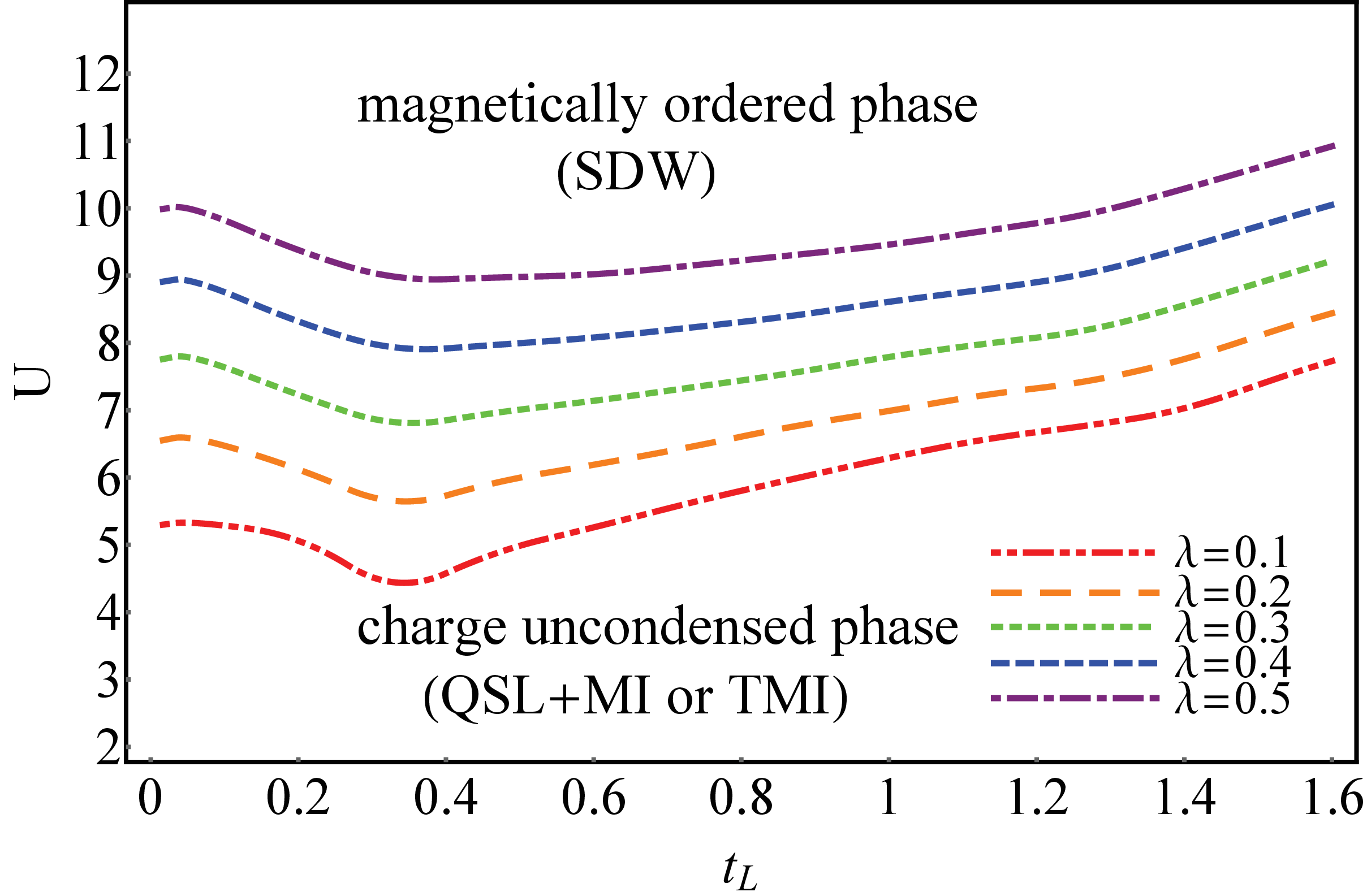}
\caption{\label{fig5}Magnetic transitions under the large-U at various strength $\lambda$ of spin-orbit coupling. SDW: spin density wave.}
\end{figure}

By applying the HF mean field decomposition, i.e. $\left<AB\right>=\left<B\right>A+\left<A\right>B-\left<A\right>\left<B\right>$, the Hubbard interaction can be obtained as
\begin{equation}
H_{U}=U\sum_{i=1}^{2N_{\Lambda}}\left[\frac{1}{4}n_{i}^{2}-\frac{1}{2}m_{i}(n_{i\uparrow}-n_{i\downarrow})+\frac{1}{4}m_{i}^{2}\right].
\label{eq36}
\end{equation}
Here $n_{i}=n_{i\uparrow}+n_{i\downarrow}$ is the total number of electrons  and $m_{i}=\left<n_{i\uparrow}-n_{i\downarrow}\right>$ is the magnetic mean field parameter at the site $i$. Considering the bipartite nature of the honeycomb lattice, we can obtain the Hubbard interaction as 
\begin{equation}
H_{U}=\frac{U}{2}\sum_{i=1}^{N_{\Lambda}}\left[-m(n_{i\uparrow}^{A}-n_{i\downarrow}^{A}-n_{i\uparrow}^{B}+n_{i\downarrow}^{B})\right]+\frac{UN_{\Lambda}}{2}m^{2}+C.
\label{eq37}
\end{equation}
Here $C=(U/4)\sum_{i=1}^{N_{\Lambda}}[(n_{i}^{A})^{2}+(n_{i}^{B})^{2}]$ is a constant and we have set $m_{i}^{A}=-m_{i}^{B}=m$ for simplicity. From the non-interacting Hamiltonian of equation~(\ref{eq2}) and the Hubbard interaction of equation~(\ref{eq37}), the total Hamiltonian of the interacting model can be written in the momentum-spin space as 
\begin{equation}
H=\sum_{\bm k}\Psi_{\bm k}^{\dagger}{\cal H}_{\bm k}\Psi_{\bm k}+\frac{UN_{\Lambda}}{2}m^{2}+C.
\label{eq38}
\end{equation}
Here the interacting Hamiltonian matrix ${\cal{H}}_{\bm{k}}$ is 
\begin{eqnarray}
\left(
\begin{array}{cccc}
\lambda \gamma-\frac{U}{2}m&-g_{t_{L}}& 0 &0\\
-g_{t_{L}}^{*}&-\lambda \gamma+\frac{U}{2}m&0&0\\
0&0&-\lambda \gamma+\frac{U}{2}m&-g_{t_{L}}\\
0&0&-g_{t_{L}}^{*}&\lambda \gamma-\frac{U}{2}m\\
\end{array}
\right).\nonumber
\end{eqnarray}
We can immediately diagaonlize the Hamiltonian matrix to obtain the free energy
\begin{equation}
F(m)=-2\sum_{\bm k}\sqrt{|g_{t_{L}}|^{2}+(\lambda\gamma-\frac{Um}{2})^{2}}+\frac{UN_{\Lambda}}{2}m^{2}+C.
\label{eq39}
\end{equation}
Then, the self-consistency equation can be obtained by minimizing the free energy as 
\begin{equation}
m=\frac{1}{N_{\Lambda}}\sum_{\bm{k}}\frac{U^{\prime}_{c}m/4-\lambda\gamma}{\sqrt{{|g_{t_{L}}|^{2}+(\lambda\gamma-U^{\prime}_{c}m/4)^{2}}}}.
\label{eq40}
\end{equation}
Here, corresponding to the replacement of $U$ by $U/2$ in the slave-rotor mean field method, we have replaced $U_{c}^{\prime}$ by $U_{c}^{\prime}/2$. The critical Hubbard interaction $U_{c}^{\prime}$ at which the magnetic transition occurs can be obtained self-consistently from equation~(\ref{eq40}) as shown in figure~\ref{fig5}.

In the large-U region, $m\not=0$ and the SDW emerges from the strong interaction. Below the boundary of the magnetic transition, the magnetic order $m$ is equal to zero and the charge uncondensed phase is obtained from the slave-rotor mean field theory. We find that $U_{c}^{\prime}$ increases with the increase of  the spin-orbit coupling $\lambda$. It can be understood from roles played by NN hopping and NNN spin-orbit coupling in the formation of the SDW phase. It is well known that two electrons on NN sites tend to couple antiferromagnetically due to the interplay between repulsive on-site Hubbard interactions and their delocalization energy. The isotropic antiferromagnetic Heisenberg exchange coupling constant (i.e. \textit{superexchange} coupling constant)\cite{1994Auerbach} $J$ is proportional to $t^{2}/U$. Similarly, the antiferromagnetic coupling will be generated between the two electrons with spin-orbit coupling on NNN sites and the superexchange coupling constant $J^{\prime}$ is proportional to $\lambda^{2}/U$. The latter competes with the antiferromagnetically ordered state on the NN sites (i.e. SDW phase). So the larger U is needed to stabilize the SDW phase when the strength of spin-orbit coupling increases. From the numerical result, we find that the minimun of critical Hubbard interaction  $U_{c}^{\prime}$ appears at $t_{L}=1/3$ for each $\lambda$. This means that the long-range hopping stabilizes the SDW phase when $t_{L}<1/3$, while destabilizes SDW phase when $t_{L}>1/3$.

\subsection{\label{sec3.3}The $t_L$-driven topological transition in the region of charge condensed phase}

In the model without strong interactions, a topological transition occurs at $t_{L}=1/3$, and is independent of the value of $\lambda$. It is interesting to know what happens to the topological transition when the interaction is introduced into the model. We investigate the $t_L$-driven topological transition of the interacting model in the two regions of charge condensed and uncondensed phases.

When charges condense, the density of the condensate of charges is $x_{0}$ ($x_{0}\not=0$) at the point $(\mathrm{i}v_{n}^0,\bm{k}^0)$. In this case, the self-consistency equations (\ref{eq29})--(\ref{eq31}) are not changed, but equations (\ref{eq28}),(\ref{eq32})--(\ref{eq34}) should be rewritten as 
\begin{equation}
\frac{1}{N_{\Lambda}}{\sum_{\bm{k}}}'\frac{\sqrt{U_{c}}}{2\sqrt{E_{-}^{X}-\mathrm{min}(E_{-}^{X})}}+x_{0}=1
\label{eq28.1}
\end{equation}
\begin{equation}
Q_{f}=\frac{1}{12N_{\Lambda}t}{\sum_{\bm{k}}}'\frac{\sqrt{U_{c}}[2Q_{X}|g|^{2}+Q_{X}^{\prime\prime}(gg_{t_{L}}^{\prime*}+g_{t_{L}}^{\prime}g^{*})]}{2\sqrt{E_{-}^{X}-\mathrm{min}(E_{-}^{X})}|Q_{X}g+Q_{X}^{\prime\prime}g_{t_{L}}^{\prime}|}+x_{0}
\label{eq32.1}
\end{equation}
\begin{eqnarray}
Q_{f}^{\prime}=\frac{1}{12N_{\Lambda}}{\sum_{\bm{k}}}'\frac{\sqrt{U_{c}}\gamma_{X}}{2\sqrt{E_{-}^{X}-\mathrm{min}(E_{-}^{X})}}+x_{0}
\label{eq33.1}
\end{eqnarray}
\begin{eqnarray}
Q_{f}^{\prime\prime}=\frac{1}{12N_{\Lambda}t_{L}}{\sum_{\bm{k}}}'\frac{\sqrt{U_{c}}[Q_{X}(gg_{t_{L}}^{\prime*}+g_{t_{L}}^{\prime}g^{*})+Q_{X}^{\prime\prime}|g_{t_{L}}^{\prime}|^{2}]}{2\sqrt{E_{-}^{X}-\mathrm{min}(E_{-}^{X})}|Q_{X}g+Q_{X}^{\prime\prime}g_{t_{L}}^{\prime}|}+x_{0}.
\label{eq34.1}
\end{eqnarray}

In the section \ref{sec3.1}, the boundary of the Mott transition for each $\lambda$ is obtained from self-consistency equations (\ref{eq28})--(\ref{eq34}). In the case of $x_{0}\not=0$, what can be obtained from the new self-consistency equations (\ref{eq29})--(\ref{eq31}), (\ref{eq28.1})--(\ref{eq34.1})? If values of $t_L$ and $U_{c}$ are chosen in the region of charge condensed phase, the mean field parameters and the density $x_{0}$ can be obtained numerically from these new self-consistency equations. It is clear that different values of $t_L$ and $U_{c}$ lead to different mean field parameters and the density $x_{0}$. If we choose all values of $t_L$ and $U_{c}$ in the region of charge condensed phase, what we do is nothing but scan the region and obtain the corresponding mean field parameters and the density $x_{0}$ for each point ($t_L$, $U_{c}$) of the region. More
interesting results may be obtained if the band structure of the spin degree of freedom is considered, since it can be topologically trivial or non-trivial. When the long-range hopping $t_L$ is small, the system stays in the QSH state as long as $t_L$ keeps the gap of the spin degree of freedom opened. If the gap closes (then reopens) for some larger $t_L$, the topological transition occurs. The situation arises in the non-interacting  $t_L$-KM model and we have found that $t_L=1/3$. What about the case of the $t_L$-KMH model? We can obtain numerically mean field solutions for all points ($t_L$, $U_{c}$) in the region of the charge condensed phase. If solutions at some points cause the gap of the spin degree of freedom to close, the set of these points is just the boundary of the topological transition which we search for. In the following, we will find the condition that the gap of the spin degree of freedom closes and then obtain the boundary of the topological transition through the self-consistency equations with the condition. 

From section \ref{sec3.1}, the single-particle energy spectrum of the spin degree of freedom (i.e. $f$-field) which may possess a topologically non-trivial band structure can be obtained as 
\begin{eqnarray}
\Xi(\bm{k})=\pm\sqrt{|Q_{f}g+Q_{f}^{\prime\prime}g_{t_{L}}^{\prime}|^{2}+\big(Q_{f}^{\prime}\lambda\gamma_{f})^{2}}.
\label{eq45}
\end{eqnarray}
On the other hand, the single-particle energy spectrum of non-interacting electrons has been obtained in equation~(\ref{eq3}). The fact that the gap of non-interacting electrons closes (then reopens) at $t_{L}=1/3t$ is critical to our investigations of the $t_L$-driven topological transition in the case of the strong interaction. By comparing equation~(\ref{eq3}) and equation~(\ref{eq45}), we find that a condition for the closure of the gap of the spin degree of freedom, i.e. the occurrence of the topological transition should be 
\begin{eqnarray}
t_{L}^{R}=\frac{1}{3}t^{R}.
\label{eq46}
\end{eqnarray}
Here $t^{R}=Q_{f}t$ and $t_{L}^{R}=Q_{f}^{\prime\prime}t_{L}$ are respectively renormalized strengths of the NN hopping and the long-range hopping. For the $t_L$-driven topological transition, the seven self-consistency equations in the charge condensed phase should be solved under the condition~(\ref{eq46}). If these equations are solvable, the gap of the spin degree of freedom must be closed and the topological transition occurs. Then, critical values of the Hubbard interaction U, the long-range hopping $t_{L}$ or the spin-orbit coupling $\lambda$ for the topological transition can be obtained from mean field solutions.
\begin{figure}
\centering
\includegraphics[width=8.2cm,height=5.1cm]{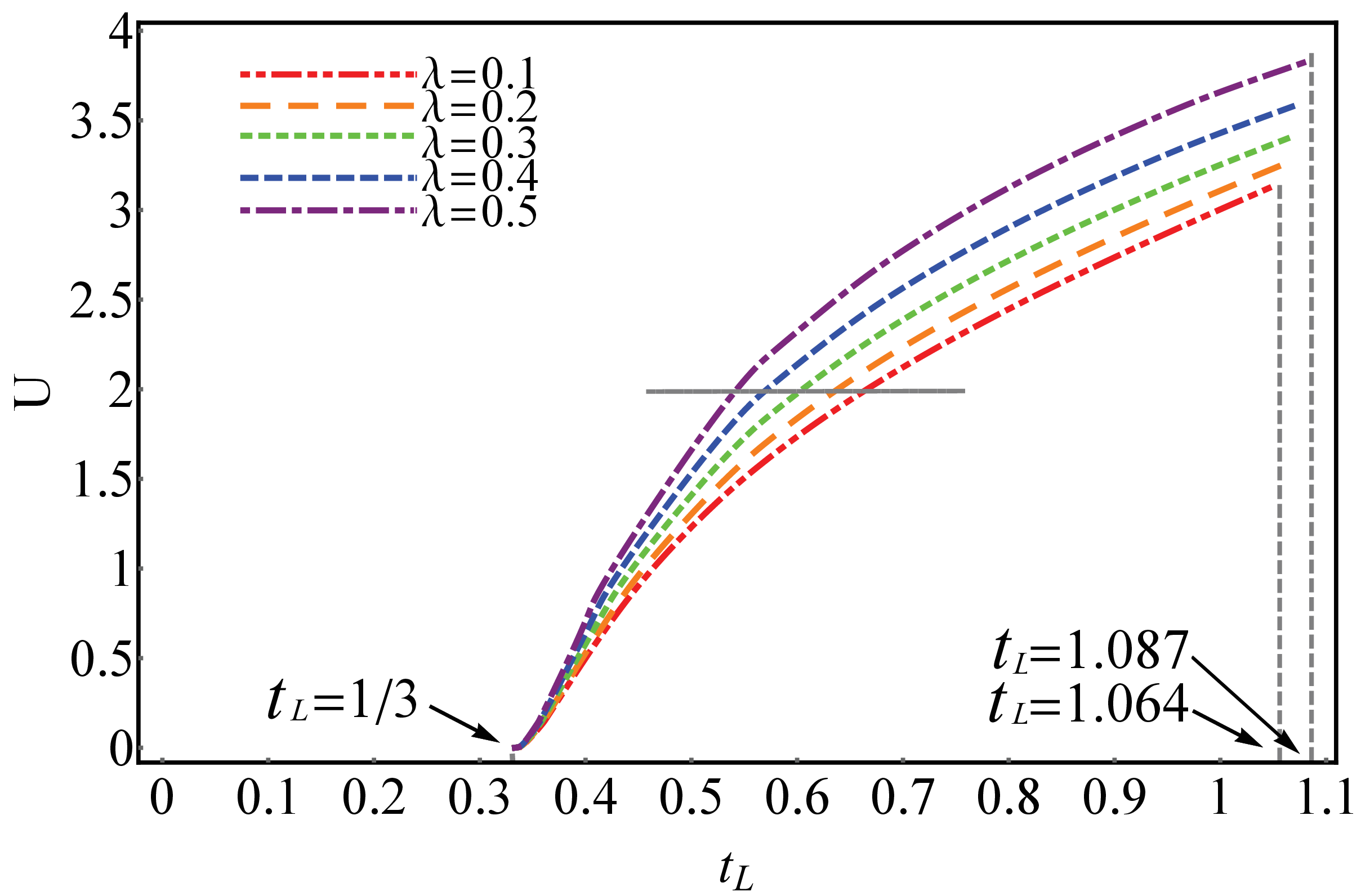}
\caption{\label{fig6}Critical curves along which gaps of the spin degree of freedom close in the region of charge condensed phase. $t_L=1.064$ is the maximum strength of the long-range hopping at which the Mott transition of the charge degree of freedom occurs for $\lambda=0.1$, while $t_L=1.078$ for $\lambda=0.5$. The horizontal solid line (gray) denotes $U=2$.}
\end{figure}
In our numerical calculation, we find that the $t_L$-driven topological transition is present in the charge condensed phase. The result is shown in figure~\ref{fig6}. The gap of spin degree of freedom closes along the critical curve. In the $U-t_{L}$ plane, the phase in the upper left region is actually the correlated QSH state discussed in the previous subsection, while a topologically trivial band insulator in the lower right region. In the region of charge condensed phase, the maximum value of $t_L$ at which the Mott transition occurs is obtained numerically, e.g. 1.064 for $\lambda=0.1$ and 1.087 for $\lambda=0.5$. The results are consistent with the calculation of mean field equations (\ref{eq28})--(\ref{eq34}) under the condition~(\ref{eq46}).

\begin{figure}
\centering
\includegraphics[width=8.2cm,height=5.1cm]{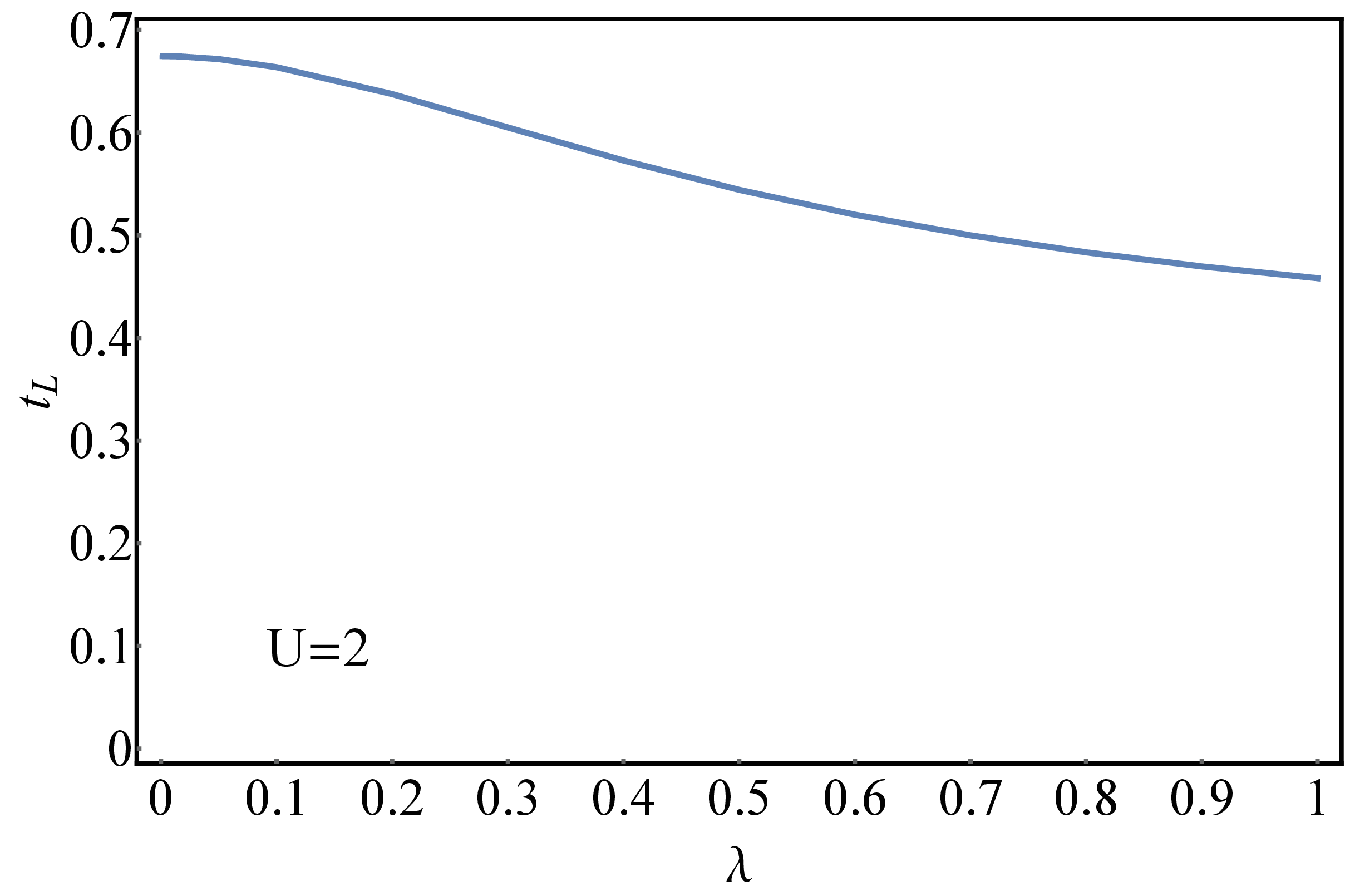}
\caption{\label{fig7}The $t_{L}-\lambda$ curve along which the topological transition occurs for $U=2$.}
\end{figure}

The first question we want to answer is whether the critical value of $t_{L}$ at which the topological transition occurs is dependent on the strength of the spin-orbit coupling when the interaction is introduced. Our calculation show that the critical value of $t_{L}$ decreases with increasing $\lambda$ due to the interaction. It is different from the case of the non-interacting limit. The relation between $t_{L}$ and $\lambda$ at $U=2$ is shown in figure~\ref{fig7}. For $U=0$ we reproduce the earlier result that the $t_{L}$-driven topological transition occurs at $t_{L}=1/3$ which is independent of the value of $\lambda$ (see figure~\ref{fig6}).

Using the QMC simulation, Hung \textit{et al.} \cite{2014Hung} observed a shift of the critical $t_{L}$ due to the introduction of interactions. We also find a shift of the critical $t_{L}$ using the slave-rotor mean field theory here. From  figure~\ref{fig6}, it is clear that the critical $t_{L}$ shifts to the larger value with inceasing $U$ for each $\lambda$. The region of correlated QSH phase is enlarged and the interaction stabilizes the correlated topological phase. The influence of interactions on the $t_{L}$-driven topological transition investigated by our mean field theory is similar to the  investigation by QMC simulation.

\begin{figure}
\centering
\includegraphics[width=8.2cm,height=5.1cm]{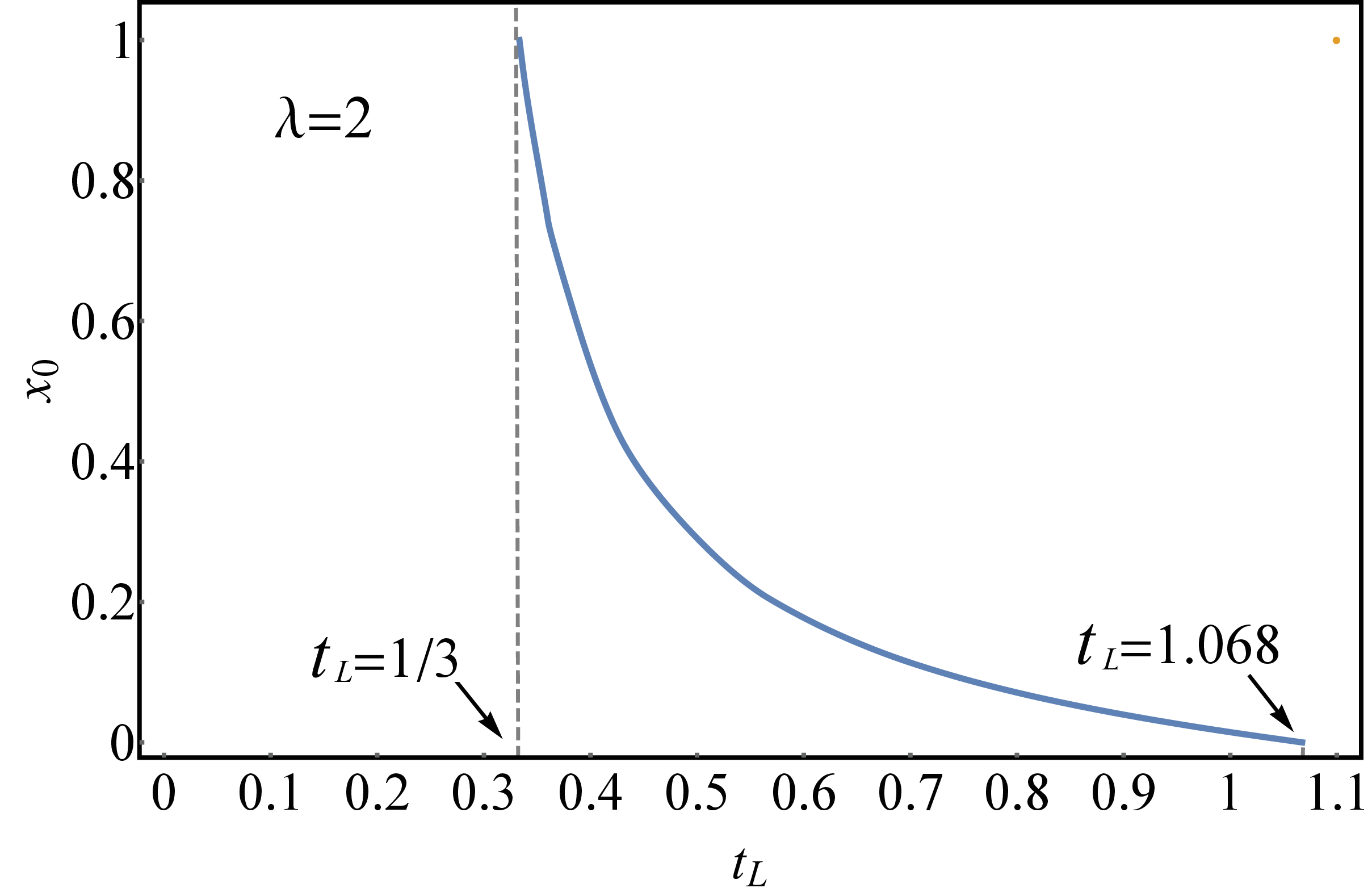}
\caption{\label{fig8}Condensate density $x_{0}$ versus critical $t_{L}$ at $\lambda=0.2$.}
\end{figure}

It is interesting to examine the fate of the condensate of charge degree of freedom along the phase boundary. The change of the condensate density $x_{0}$ with the variation of critical $t_{L}$ is shown in figure~\ref{fig8}. For $\lambda=0.2$, the condensate density has the maximum value $x_{0}=1$ at $t_{L}=1/3$ and equals zero at $t_{L}$=1.068. The fact that the condensate density decreases with increasing critical $t_{L}$ can be understood from the relation between critical $U$ and $t_{L}$. At $t_{L}=1/3$, the Hubbard interaction $U=0$ and the condensate density $x_{0}$ should have its maximum value. The condensate density $x_{0}$ should decrease with the increase of critical $t_{L}$ due to the increasing strong interaction with $t_{L}$. At $t_L=1.068$, the condensate density equals zero because of the occurrence of the Mott transition. In our numerical calculation, the condensate density shifts slightly by roughly $0.7\%$ when the spin-orbit coupling $\lambda$ changes by 0.1. So we just draw the curve of condensate density at $\lambda=0.2$.

\subsection{\label{sec3.4}The $t_L$-driven topological transition in the Mott region}
\begin{figure}
\centering
\includegraphics[width=8.2cm,height=5.1cm]{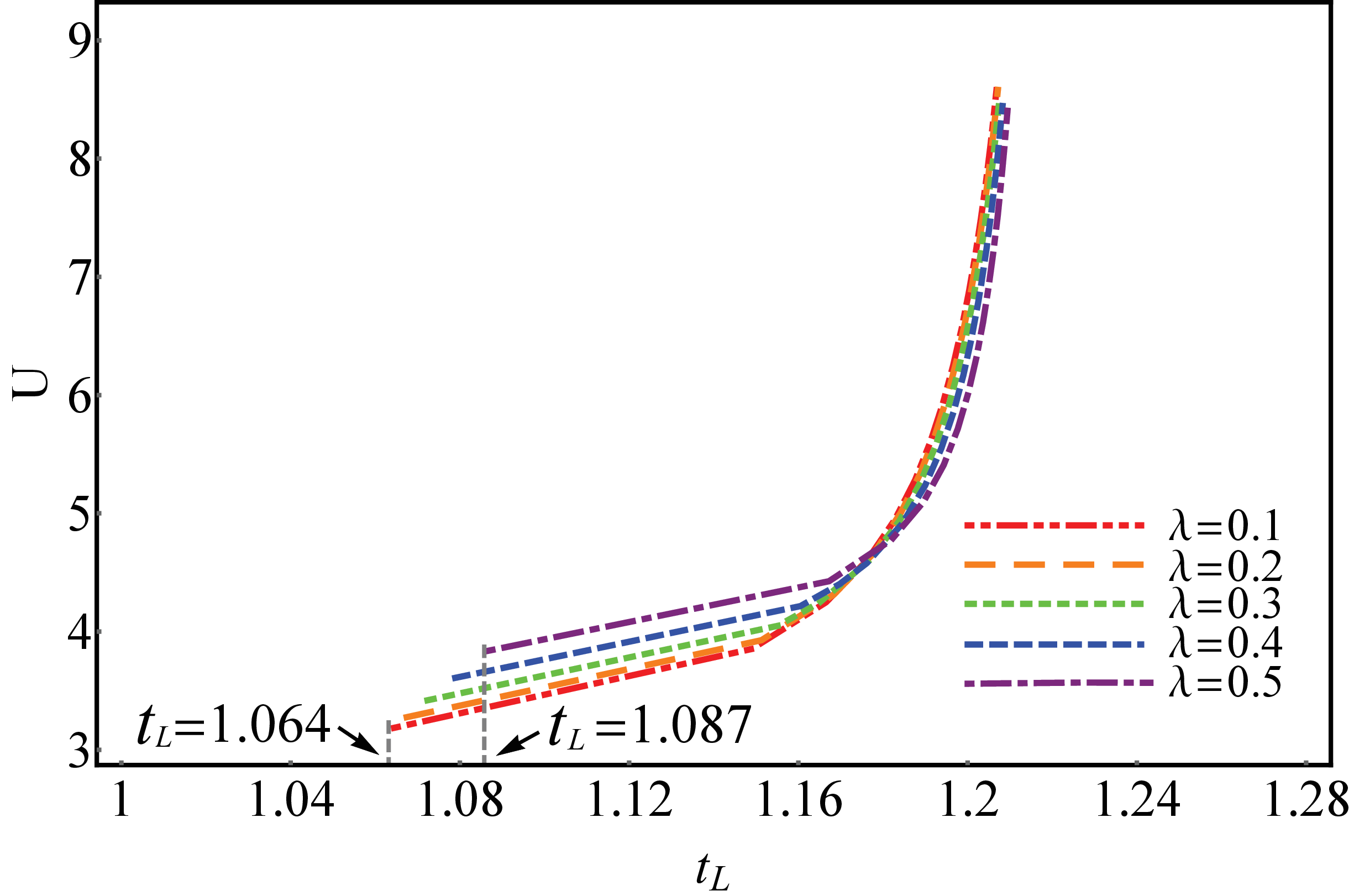}
\caption{\label{fig9}Critical curves along which gaps of spin degree of freedom close in the Mott region for various $\lambda$.}
\end{figure}
In the Mott region of the intermediate Hubbard interaction, there is no condensate of the charge degree of freedom, i.e. $x_0=0$ and $\rho\not=-\mathrm{min}(E_{-}^{X})$.  So terms $-\mathrm{min}(E_{-}^{X})$ in the self-consistency mean field equations~(\ref{eq28})-(\ref{eq34}) should be replaced by $\rho$. The situation is similar to the one discussed in section \ref{sec3.3}. Using the altered self-consistency equations, we can scan the Mott region and obtain the mean field parameters and $\rho$ for all points ($t_L$, $U_{c}$) in the Mott region. If one wants to obtain the possible topological transition in this region, the gap of the spin degree of freedom should be considered additionally. The condition~(\ref{eq46}) for the $t_L$-topological transition is already obtaind in the previous section. Then, If the altered self-consistency equations can be solved under the condition, points ($t_L$, $U_{c}$) which form the boundary of topological transition are picked out. Curves of the topological transition for various $\lambda$ are shown in figure~\ref{fig9}.
The phase in the upper left region is actually a TMI. In the lower right region, the band structure of the spin degree of freedom is topologically trivial and the phase is a mixing state of a conventional QSL state and the MI. In the Mott region, the minimum of $t_L$, e.g. 1.064 for $\lambda=0.1$ and 1.087 for $\lambda=0.5$, is the critical point at which the Mott transition of charge degree of freedom occurs. The results are consistent with the observation in the previous subsection. As the case of condensed charges, the critical value of $t_L$ at which topological transition occurs is influenced by the spin-orbit coupling and Hubbard interactions. However, there is another character about the influence, i.e. the critical $t_L$ is insensitive to the spin-orbit coupling and Hubbard interactions in the case of the larger U.
 
\begin{figure}
\centering
\includegraphics[width=8.2cm,height=5.1cm]{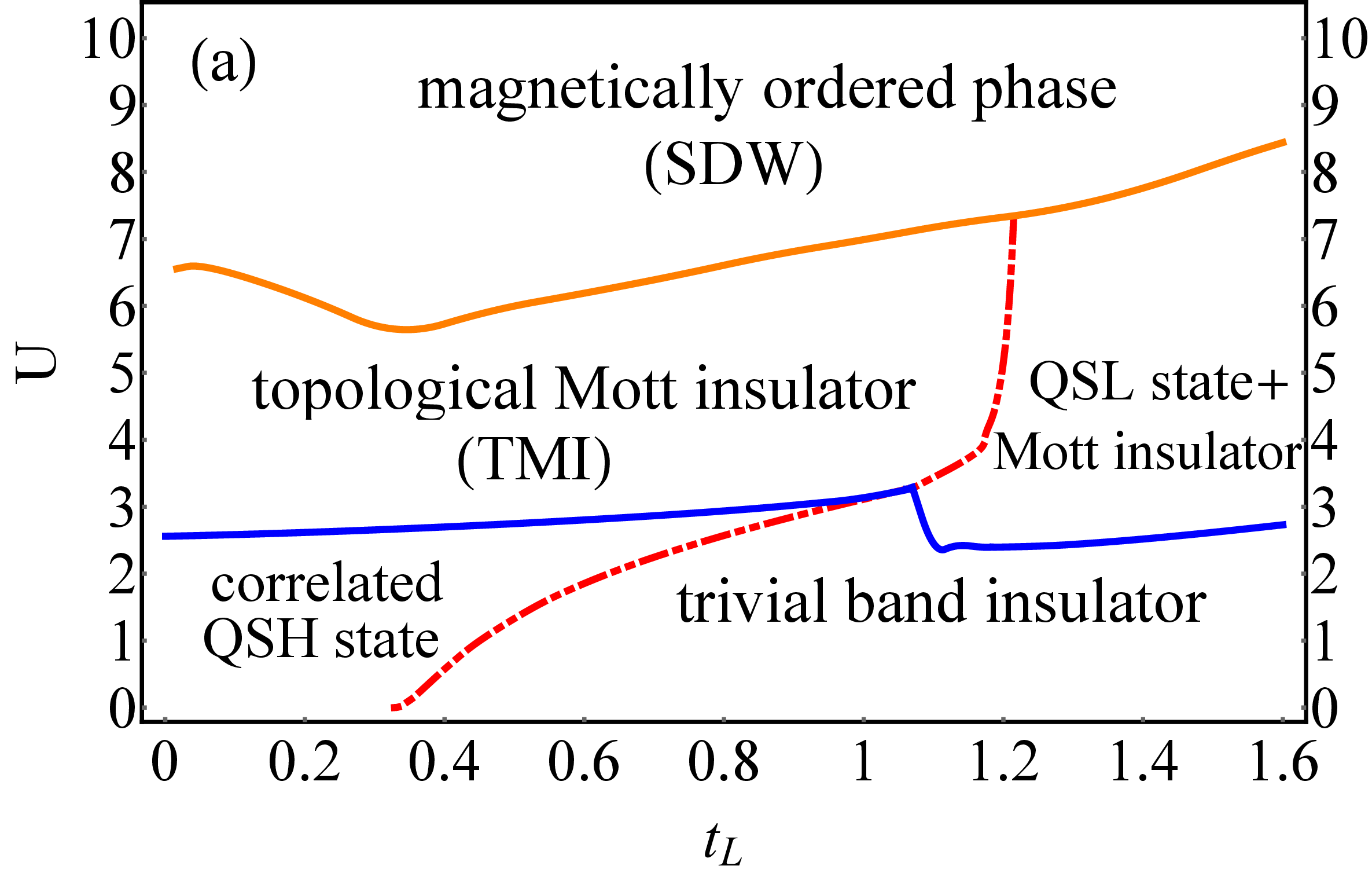}
\includegraphics[width=8.2cm,height=5.1cm]{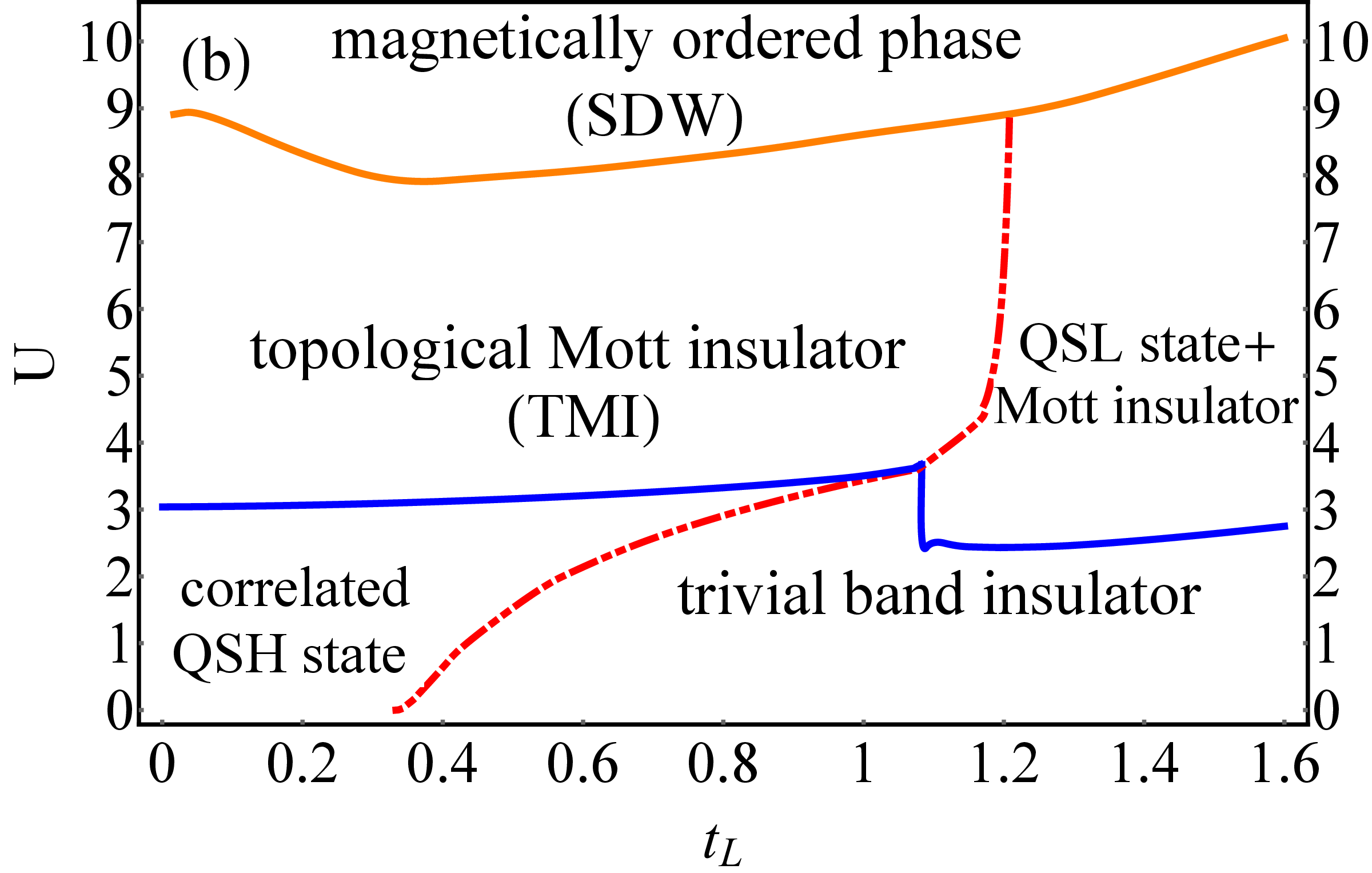}
\caption{\label{fig10}The complete phase diagram of the $t_L$-KMH model at $\lambda=0.2$ (a) and $\lambda=0.4$ (b).}
\end{figure}

We can summarize above discussions to obtain phase diagrams of the interacting topological model as shown in figure~\ref{fig10}. Differences between phase diagrams at different spin-orbit coupling $\lambda$ are quantitative. Phase boundaries for different $\lambda$ in phase diagrams can be observed in detail from  figure~\ref{fig4},~\ref{fig5},~\ref{fig6} and~\ref{fig9}. Under the Mott transition (indicated by blue solid lines in phase diagrams), the charge degree of freedom combines the spin degree of freedom to form the conventional electron. In this region, using the slave-rotor mean field theory, we recover the $t_L$-driven topological transition (indicated by red dot lines in the lower part of phase diagrams) which have been investigated by Hung \textit{et al.} \cite{2014Hung} using the QMC simulation. Our results also demonstrate that influences of interactions are ``positive", i.e. interactions can stabilize the QSH state against the long-range hopping which drives the model into topologically trivial band insulator. In the intermediate part of phase diagrams, the charge degree of freedom forms the Mott insulator and the spin degree of freedom forms a QSL state which determines the topology of the model. The part of phase diagrams, which includes the TMI, the topologically trivial mixing phase and the $t_L$-driven topological transition (also indicated by red dot lines) between the two phases, is investigated for the first time in this work. In the upper part of phase diagrams, the magnetically ordered phase of $t_L$-KMH model is also obtained for the first time, although the approach used here is the straightforward HF mean field theory. 

\section{\label{sec4}Conclusions and outlook}

There is a $t_L$-driven topological transition when the four-lattice-constant range hopping is introduced into the KM model. The critical strength of the long-range hopping at which the topological transition occurs is independent of the spin-orbit coupling $\lambda$, i.e. $t_L\equiv1/3$. We apply the slave-rotor mean field method to the interacting KM model with the long-range hopping. In the mean field method, the interaction lead to a spin-charge separation and the Mott transition of charge degree of freedom occurs at the intermediate Hubbard interaction. It is a mixing phase of a Mott insulator of charge degree of freedom and a QSL state of spin degree of freedom above the Mott transition, while a state of conventional electrons below the Mott transition. At the small long-range hopping, the charge condensed phase is a correlated QSH state which connects adiabatically to the QSH state possessed by the $t_L$-KM model, and the charge uncondensed phase is the TMI. By comparing the band of spin degree of freedom of the $t_L$-KMH model with the one of electrons in the non-interacting limit, a key condition in terms of the renormalized long-range hopping and nearest neighbor hopping (i.e. $t_{L}^{R}=1/3t^{R}$) is obtained to investigate the $t_L$-driven topological transition in regions below and above the Mott transition. Under the transition condition, the critical $t_L$ of the topological transition is got from the self-consistency mean field equations in both regions. In the region below the Mott transition, the critical $t_L$ is influenced by the strong interaction. Our result shows that the critical $t_{L}$ shifts to the larger value with inceasing $U$ for each $\lambda$. The shifts demonstrate that the strong interaction stabilizes the topological phase against the long-range hopping which drives the system from the correlated QSH insulator to the topologically trivial band insulator. The investigation is  qualitatively consistent with the QMC simulation, although the slave-rotor mean field treatment overestimates the effect of the strong interaction on the $t_L$-driven topological transition. Moreover, from arguments of Florens \textit{et al.} \cite{2002Florens,2003Florens,2004Florens} this mean field approach is still reliable in the intermediate interaction and can provide some reasonable physics of strongly correlated systems. In the region above the Mott transition, a $t_L$-driven topological transition of the spin degree of freedom is obtained for the first time and we find that it is insensitive to the spin-orbit coupling and the Hubbard interaction when the Hubbard interaction U becomes large in the Mott region. Adding the magnetic transition from the HF mean field method, we obtain the complete phase diagram of the $t_L$-KMH model, which displays the rich physics stemming from the interplay of topology, electron correlations and the long-range hopping.

There are many generalized models possessing topological states, e.g. KM model with staggered potentials or third-nearest neighbor hopping. It is well known that there are topological transitions driven by the staggered potential or others in these models. Influences of strong interaction on topological transitions in these models may be investigated by the slave-rotor mean field theory, since the mean field theory can reveal appropriately some physics of strongly correlated systems. Moreover, if lattice symmetries are considered, the investigation may be more interesting.

\ack
This work was supported by Applied Basic Research Program of Yunnan Provincial Science and Technology Department (No. 2013FZ083) and Foundation of Yunnan Provincial Education Department (No. 2012Z040). TD also acknowledge the support from NSFC under grant Nos. 11464051, 61640415.


\section*{References}
\begin{thebibliography}{<42>}
\bibitem {2010Hasan}Hasan M Z and Kane C L 2010 {\it Rev. Mod. Phys.} 82 3045
\bibitem {2011Qi}Qi X L and Zhang S C 2011 {\it Rev. Mod. Phys.} 83 1075
\bibitem {2013Ando}Ando Y J 2013 {\it Phys. Soc. Jpn.} 82 102001
\bibitem {1982Thouless}Thouless D J, Kohmoto M, Nightingale M P and den Nijs M 1982 {\it Phys. Rev. Lett.} 49 405
\bibitem {1988Haldane}Haldane F D M 1988 {\it Phys. Rev. Lett.} 61 2015
\bibitem {2005aKane}Kane C L and Mele E J 2005 {\it Phys. Rev. Lett.} 95 146802
\bibitem {1984Niu}Niu Q and Thouless D J 1984 {\it J. Phys. A: Math. Gen.} 17 2453
\bibitem {1985Niu}Niu Q, Thouless D J and Wu Y S 1985 {\it Phys. Rev. } B 31 3372
\bibitem {2010Pesin}Pesin D A and Balents L 2010 {\it Nat. Phys.} 6 376
\bibitem {2015Neupert}Neupert T, Chamon C, Ladecola  T, Santos L H and Mudry C 2015 {\it Phys. Scr.} T164 014005
\bibitem {2015Maciejko}Maciejko J and Fiete G A 2015 {\it Nat. Phys.} 11 385
\bibitem {2013Hohenadler}Hohenadler M and  Assaad F F 2013 {\it J. Phys.: Condens. Matter} 25 143201
\bibitem {2014Imada}Imada M, Yamaji Y, and Kurita M 2014 {\it J. Phys. Soc. Jpn.} 83 061017
\bibitem {2014Witczak-Krempa}Witczak-Krempa W, Chen G, Kim Y B and Balents L 2014 {\it Ann. Rev. Condens. Matter Phys.} 5  57
\bibitem {2005bKane}Kane C L and Mele E J 2005 {\it Phys. Rev. Lett.} 95 226801
\bibitem {1963Hubbard}Hubbard J 1963 {\it Proc. R. Soc.} A 276 238
\bibitem {1964Hubbard}Hubbard J 1964 {\it Proc. R. Soc.} A 281 401
\bibitem {1998Imada}Imada M, Fujimori A, and Tokura Y 1998 {\it Rev. Mod. Phys.} 70 1039
\bibitem {2008Young}Young M W, Lee S S and Kallin C 2008 {\it Phys. Rev.} B 78 125316
\bibitem {2010Rachel}Rachel S and Le Hur K 2010 {\it Phys. Rev.} B 82 075106
\bibitem {2012Ruegg}R\"uegg A and Fiete G A 2012 {\it Phys. Rev. Lett.} 108 046401
\bibitem {2012Vaezi}Vaezi A, Mashkoori M and Hosseini M 2012 {\it Phys. Rev.} B 85 195126
\bibitem {2012Wu}Wu W, Rachel S, Liu W M and Le Hur K 2012 {\it Phys. Rev.} B 85 205102
\bibitem {2011Yu}Yu S L, Xie X C and Li J X 2011 {\it Phys. Rev. Lett.} 107 010401
\bibitem {2011Hohenadler}Hohenadler M, Lang T C and Assaad F F 2011 {\it Phys. Rev. Lett.} 106 100403 
\bibitem {2011Zheng}Zheng D, Zhang G M and Wu C 2011 {\it Phys. Rev.} B 84 205121
\bibitem {2012Hohenadler}Hohenadler M, Meng Z Y, Lang T C, Wessel S, Muramatsu A  and Assaad F F 2012  {\it Phys. Rev.} B 85 115132
\bibitem {2012Griset}Griset C and Xu C 2012 {\it Phys. Rev.} B 85 045123 
\bibitem {2014Bercx}Bercx M, Hohenadler M and Assaad F F 2014 {\it Phys. Rev.} B 90 075140
\bibitem {2018Rachel}Rachel S 2018 (arXiv:1804.10656)
\bibitem {2013Hung}Hung H H, Wang L, Gu Z C and Fiete G A 2013 {\it Phys. Rev.} B 87 121113(R)
\bibitem {2014Hung}Hung H H, Chua V, Wang L and Fiete G A 2014 {\it Phys. Rev.} B 89 235104
\bibitem {2015Chen}Chen Y H, Hung H H, Su G, Fiete G A and Ting C S 2015 {\it Phys. Rev.} B 91 045122 
\bibitem {2013Lang}Lang T C, Essin A M, Gurarie V and Wessel S 2013 {\it Phys. Rev.} B 87 205101 
\bibitem {2014Lai and Hung}Lai H H and Hung H H 2014 {\it Phys. Rev.} B 89 165135
\bibitem {2002Florens}Florens S and Georges A 2002 {\it Phys. Rev.} B 66 165111
\bibitem {2003Florens}Florens S, San Jos\'e P, Guinea F and  Georges A 2003 {\it Phys. Rev.} B 68  245311
\bibitem {2004Florens}Florens S and Georges A 2004 {\it Phys. Rev.} B 70 035114 
\bibitem {2005Lee}Lee S S and Lee P A 2005 {\it Phys. Rev. Letts.} 95 036403
\bibitem {2007Zhao}Zhao E and Paramekanti A 2007 {\it Phys. Rev.} B 76 195101
\bibitem {2008Senthil}Senthil T 2008 {\it Phys. Rev.} B 78 045109
\bibitem {1996Sorella}Sorella S and Tosatti E 1992 {\it Europhys. Lett.} 19 699
\bibitem {2009Jafari}Jafari S A 2009 {\it Eur. Phys. J.} B 68 537
\bibitem {2015Coleman}Coleman P 2015 \textit{Introduction to Many-Body Physics} (Cambridge: Cambridge University Press)
\bibitem {2011He}He J, Kou S P, Liang Y and Feng S 2011 {\it Phys. Rev.} B 83 205116
\bibitem {2016Li}Li K, Yu S L, Gu Z L and Li J X 2016 {\it Phys. Rev.} B 94 125120
\bibitem {1994Auerbach}Auerbach A 1994 \textit{Interacting electrons and quantum magnetism} (New York: Springer-Verlag)
\end{thebibliography}
\end{document}